\documentclass[twocolumn,twocolappendix]{aastex63}
\usepackage{amsmath}

\newcommand\kms{km s$^{-1}$}

\newcommand\teff{$T_{\rm eff}$}
\newcommand\logteff{$\log T_{\rm eff}$}
\newcommand\logg{$\log g$}

\defcitealias{kounkel2018a}{Paper I}
\usepackage{listings}

\begin{document}
\title{Gaia Net: Towards robust spectroscopic parameters of stars of all evolutionary stages}

\author{Dylan Huson}
\affil{Computer Science Department, Western Washington University, 516 High Street, Bellingham, WA, 98225, USA, USA}
\author{Indiana Cowan}
\affil{Computer Science Department, Western Washington University, 516 High Street, Bellingham, WA, 98225, USA, USA}
\author{Logan Sizemore}
\affil{Computer Science Department, Western Washington University, 516 High Street, Bellingham, WA, 98225, USA, USA}
\author[0000-0002-5365-1267]{Marina Kounkel}
\affil{Department of Physics and Astronomy, University of North Florida, 1 UNF Dr, Jacksonville, FL, 32224, USA}
\email{marina.kounkel@unf.edu}

\author[0000-0002-5537-008X]
{Brian Hutchinson}
\affil{Computer Science Department, Western Washington University, 516 High Street, Bellingham, WA, 98225, USA, USA}
\affil{Human-Earth Systems \& Science Group, Pacific Northwest National Laboratory, 5825 University Research Court, Suite 3500, College Park, MD, 20740, USA}
\email{Brian.Hutchinson@wwu.edu}

\begin{abstract}

We present a new processing of XP spectra for 220 million stars released in Gaia DR3. The new data model is capable of handling objects with \teff\ between 2000 and 50,000 K, and with \logg\ between 0 and 10, including objects of multitude of masses and evolutionary stages. This includes for the first time ever robust processing of spectroscopic parameters for pre-main sequence stars, with \logg\ sensitivity towards their age. Through this analysis we examine the distribution of young low mass stars with ages of up to 20 Myr in the solar neighborhood, and we identify a new massive ($>1000$ stars) population, Ophion, which is found east of Sco Cen. This population appears to be fully disrupted, with negligible kinematic coherence. Nonetheless, due its young age it appears to still persist as a spacial overdensity. Through improved determination of ages of the nearby stars, it may be possible to better recover star forming history of the solar neighborhood outside of the moving groups.
\end{abstract}


\section{Introduction}

Since its first data release, Gaia has revolutionized galactic astronomy. In addition to providing an unprecedented precision in its astrometry, Gaia has provided some of the best photometry for stars \citep{gaia-collaboration2018,gaia-collaboration2018a,gaia-collaboration2023}. Indeed, in the recent years, G, BP, and RP fluxes appear to have surpassed Johnson filters in the widespread of use due to a homogeneous coverage across the entire sky and great sensitivity.

Gaia also has performed the largest spectroscopic survey to date. Not only has it obtained medium resolution (RVS) spectra for 33 million stars, it also has also produced low resolution (XP) spectra for hundreds of millions of stars. \citep{andrae2023,fouesneau2023,recio-blanco2023,de-angeli2023}. Furthermore, BP/RP fluxes are calculated from integrated XP spectra \citep{riello2021}.

Gaia has processed these data to provide a number of stellar parameters, such as \teff, \logg, and [Fe/H]. However, as they have relied on fitting the data through using synthetic models, the extracted stellar parameters were particularly poor for sources that were found outside of the region covered by these synthetic models, or in regions where the models are poorly calibrated to the real data.

There have been a few different attempts at re-reducing these spectra. For example, \citet{zhang2023} have applied machine learning techniques to derive \teff, \logg, and [Fe/H] from scratch. Specifically, they have trained a feed-forward model using stellar parameters from Large Sky Area Multi-Object Fiber Spectroscopic Telescope (LAMOST) spectra. This model is capable of generating an XP model spectrum for given input stellar parameters, which is then capable of estimating parameters for the XP spectra. LAMOST is a large spectroscopic survey, having to-date obtained medium resolution (R$\sim$1800) spectra of more than 10 million stars. Being higher resolution than XP, it has been much more amendable to stellar parameter determination using traditional techniques. Thus, \citet{zhang2023} have identified the sources that have been observed by both surveys, assumed labels from LAMOST, applied it to the corresponding XP spectra, and trained a model that would generalize and extrapolate across the entire applicable XP dataset. The dataset consisted of the flux extracted from the coefficients, as well as broad-band 2MASS and WISE photometery.

While such an approach of label transfer is fundamental and has a lot of promise, \citet{zhang2023} had labels that were not entirely suited for such a task. The labels spanned OBAFGK range, leaving out low mass stars, pre-main sequence stars, and anything that was found below the main sequence, such as hot subdwarfs and white dwarfs. Given that Gaia XP spectra contain a variety of different type of sources, applying a model on the sources outside of the range on which it was trained on produces a number of artefacts. 

More recently, \citet{khalatyan2024} have trained a new model using a significantly greater diversity in its labels from different spectroscopic surveys, including APOGEE, GALAH, Gaia-ESO, LAMOST, RAVE, and SEGUE, supplemented with some sources to cover more unique areas of the parameters, such as very metal poor stars, hot subdwarfs, and white dwarfs. The model was trained on the coefficients used to characterize XP spectra, astrometric parameters (including position on the sky, parallax, and proper motions), as well as broad-band photometry. While the resulting catalog is extremely robust for the vast majority of all stars, it nonetheless has some limitations with regards to lower mass stars, particularly those that are pre-main sequence. 

In this work, we develop a new model, Gaia Net, that is based on our previous efforts to derive comprehensive stellar parameters for spectra obtained with APOGEE, BOSS, and LAMOST \citep{olney2020,sprague2022,sizemore2024} with the goal of providing as comprehensive of a catalog as possible of all the corners of the parameter space, with particular attention given to the pre-main sequence stars. These were the first (and to date the only) spectroscopic catalogs that provided sufficiently robust \logg s for young stars to enable their use as a proxy for age.

\section{Data}
\begin{figure}
\epsscale{1.2}
\plotone{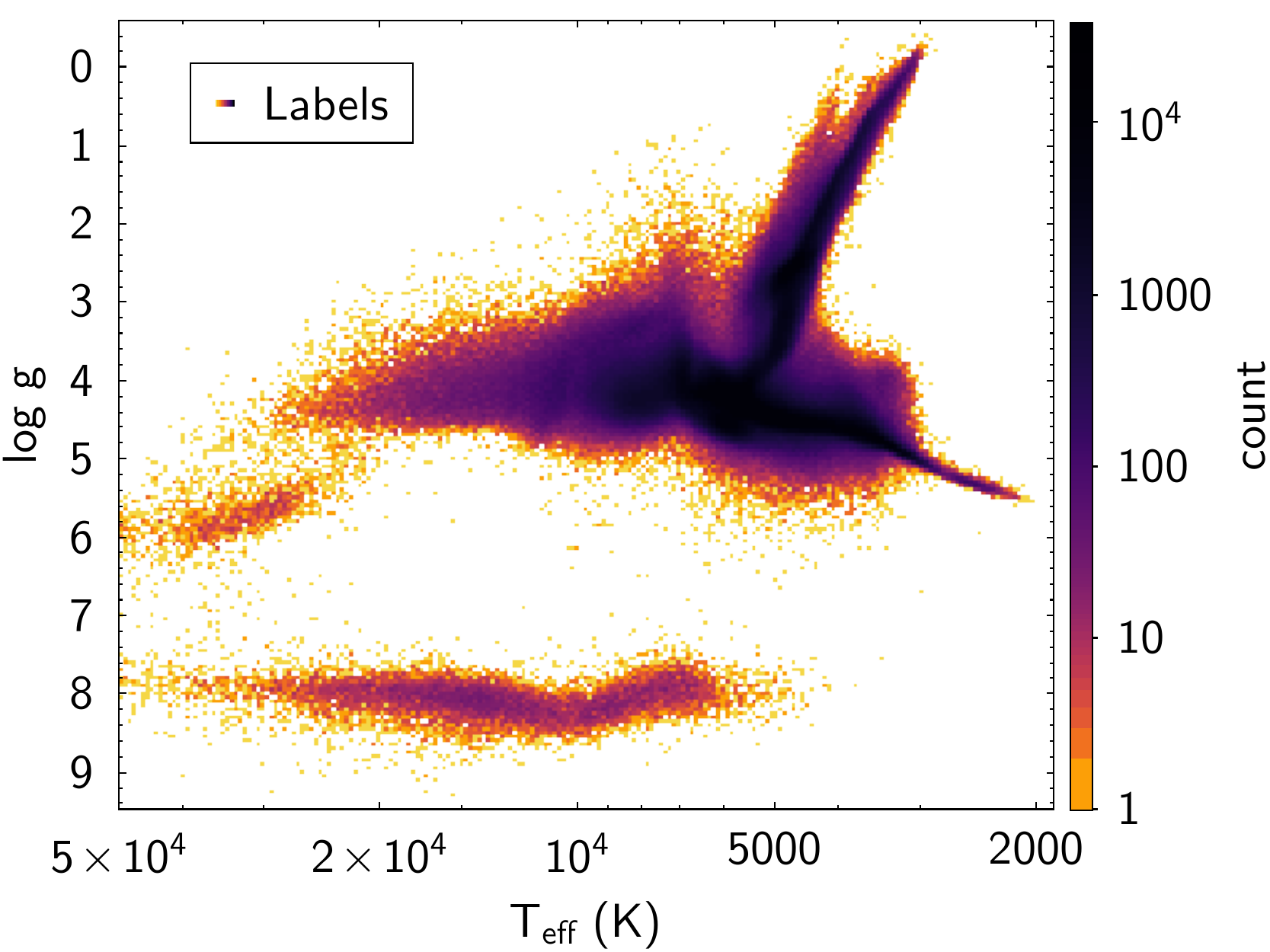}
\caption{Distribution of \teff\ and \logg\ parameters in the training sample that have been inherited from the spectra previously observed with LAMOST or with BOSS \citep{sizemore2024}
\label{fig:training}}
\end{figure}

\begin{figure}
\epsscale{1.2}
\plotone{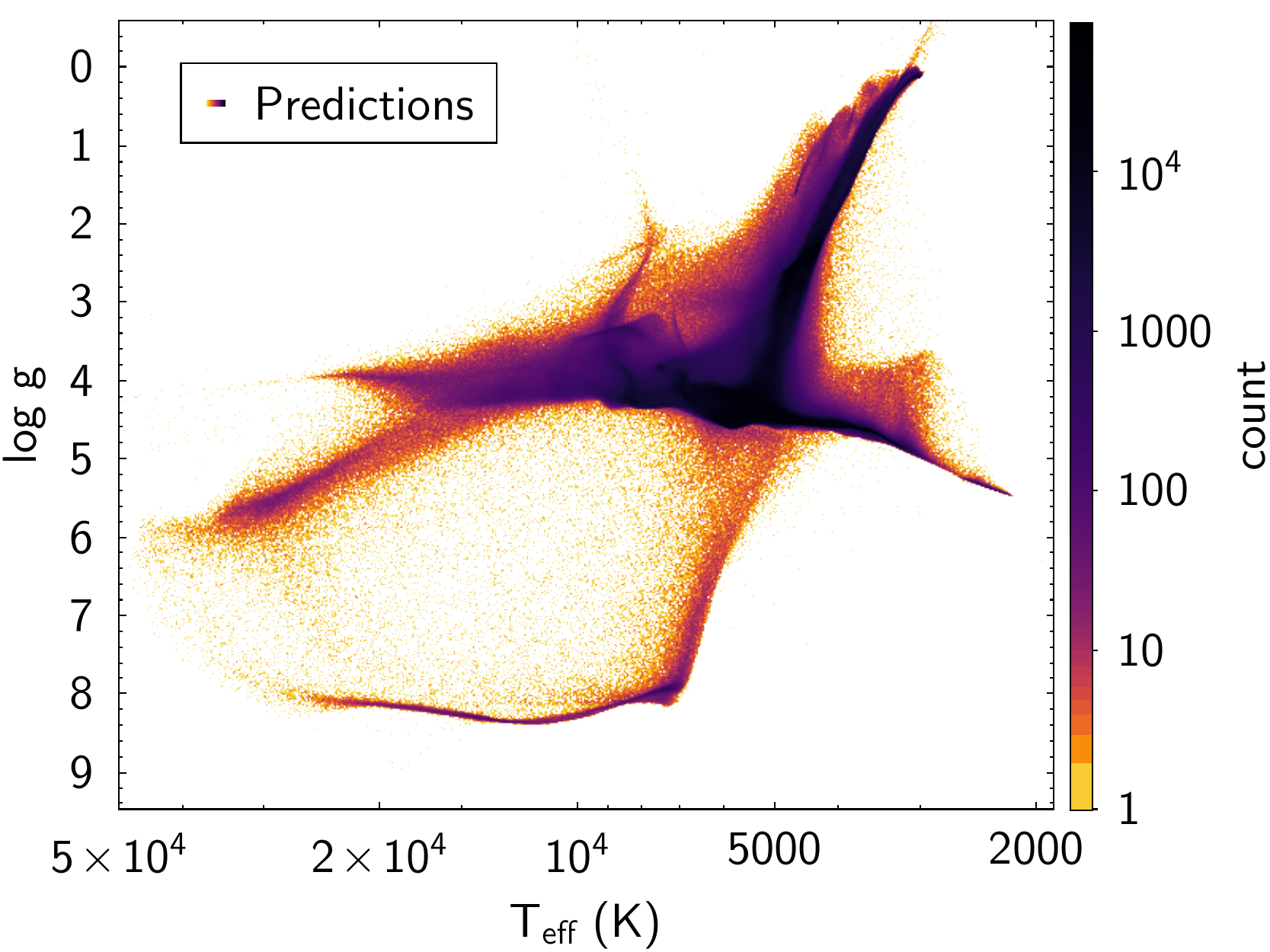}
\caption{Distribution of \teff\ and \logg\ of all of the sources evaluated by BOSS Net. 
\label{fig:predictions}}
\end{figure}

Gaia DR3 release has made available XP spectra for $\sim$220 million sources. These spectra consist of the blue BP part, spanning the wavelength range from 3300 to 6800 \AA, and the RP part, spanning the range from 6400 to 1050 \AA \citep{montegriffo2023}. For the vast majority of the sources, instead of the fluxes, these spectra are stored only as coefficients to a function needed to represent the flux, adding up to a total of 110 elements combined for both BP and RP.

 We utilize these coefficients in training the model. Unlike previous efforts by \citet{zhang2023} or \citet{khalatyan2024} to derive stellar parameters, we do not supplement these coefficients with any metadata. This includes omitting integrated photometry that could potentially help it optimize accounting for the extinction (and thus be able to more reliably \teff\ from the overall shape of the spectral energy distribution). It also excludes astrometric parameters, preventing the model from implicitly computing luminosity from the flux (to ensure that it is not capable of using luminosity as a proxy of \logg). This forces the model to learn the characteristic spectral features, however low resolution these features might be, without searching for potential shortcuts.

 Our training was constructed out of the sources that have been observed with LAMOST or with SDSS, spectral parameters of which have been derived with BOSS Net \citep{sizemore2024}. In total, these labels have been available for $\sim$5.8 million stars (Figure \ref{fig:training}). \teff\ and \logg\ are available for all stars, and [Fe/H] is assumed valid for sources with \logg$<$5 and \teff$>$3200 K.

\begin{figure}
\epsscale{1.2}
\plotone{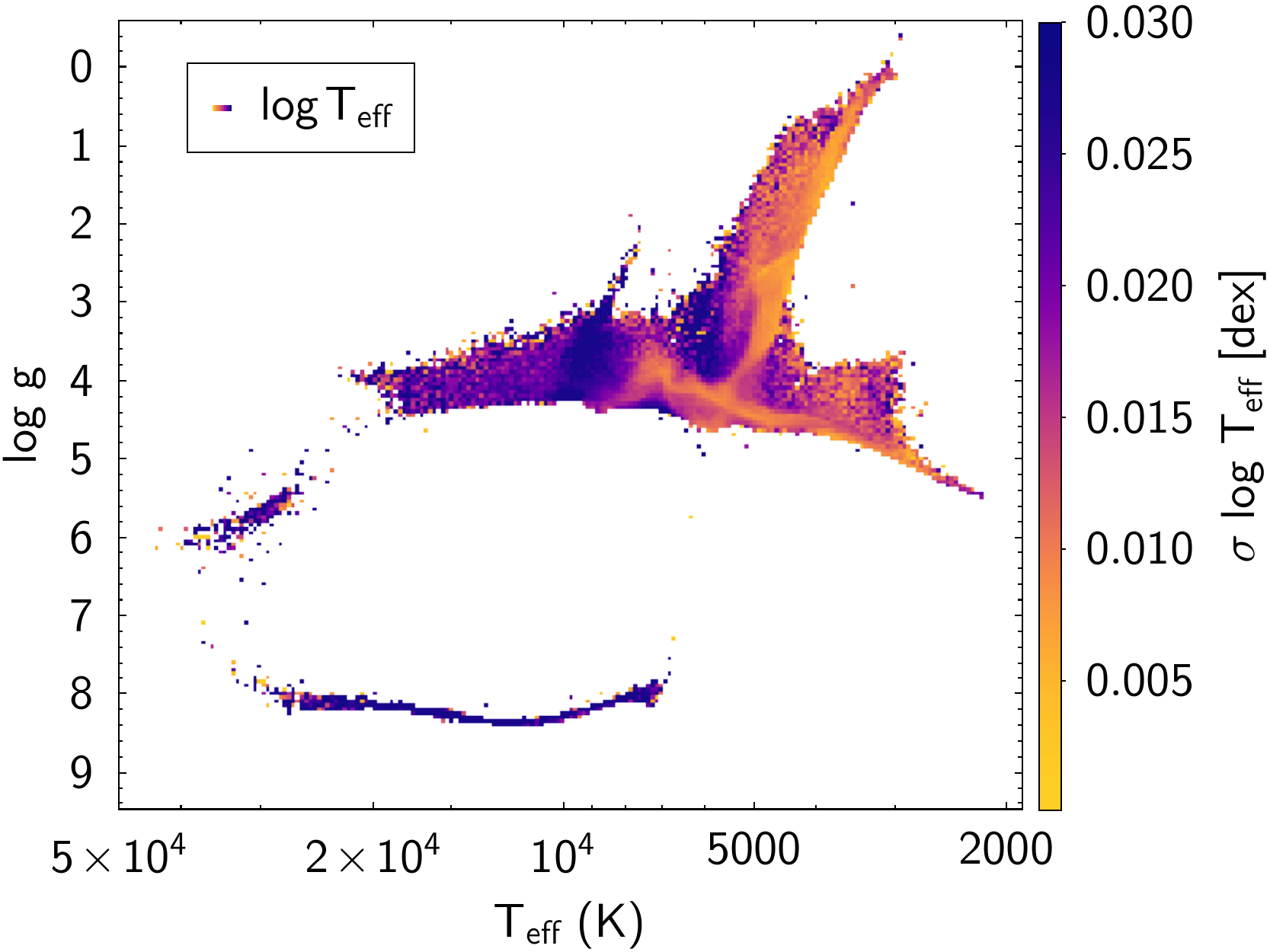}
\plotone{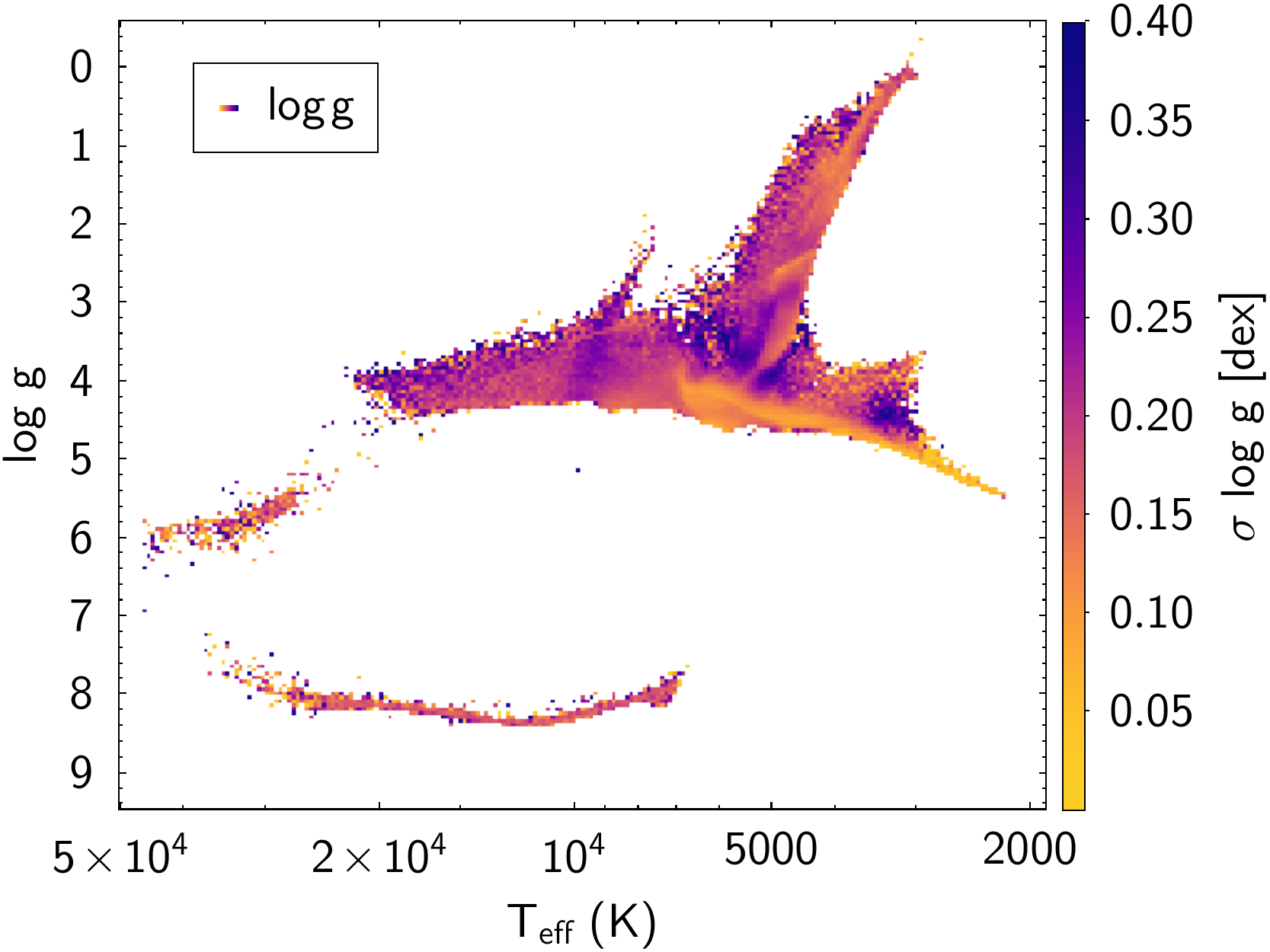}
\plotone{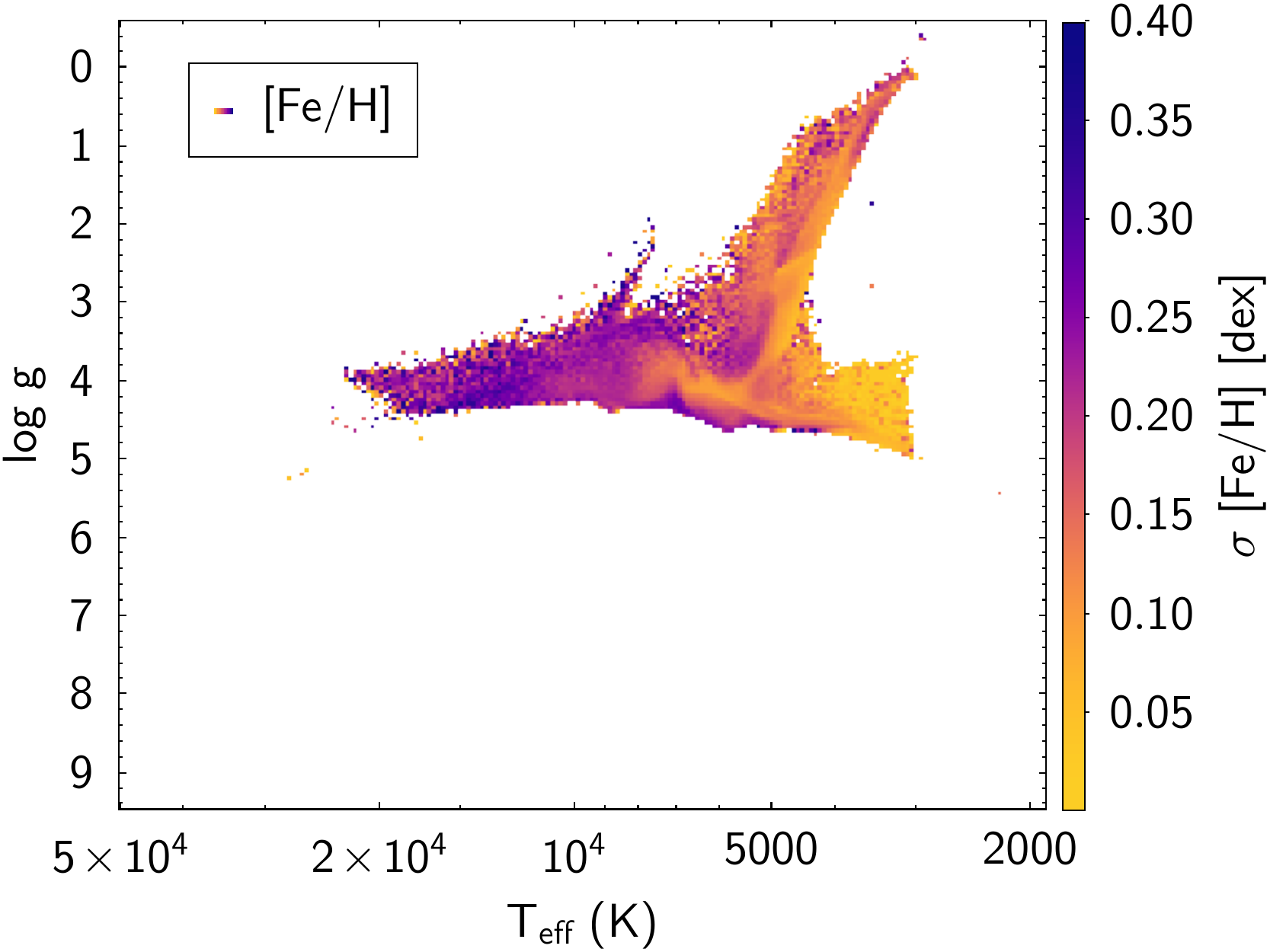}
\caption{Scatter between the predictions obtained for Gaia Net and the original labels.
\label{fig:errors}}
\end{figure}

\begin{figure}
\epsscale{1.2}
\plotone{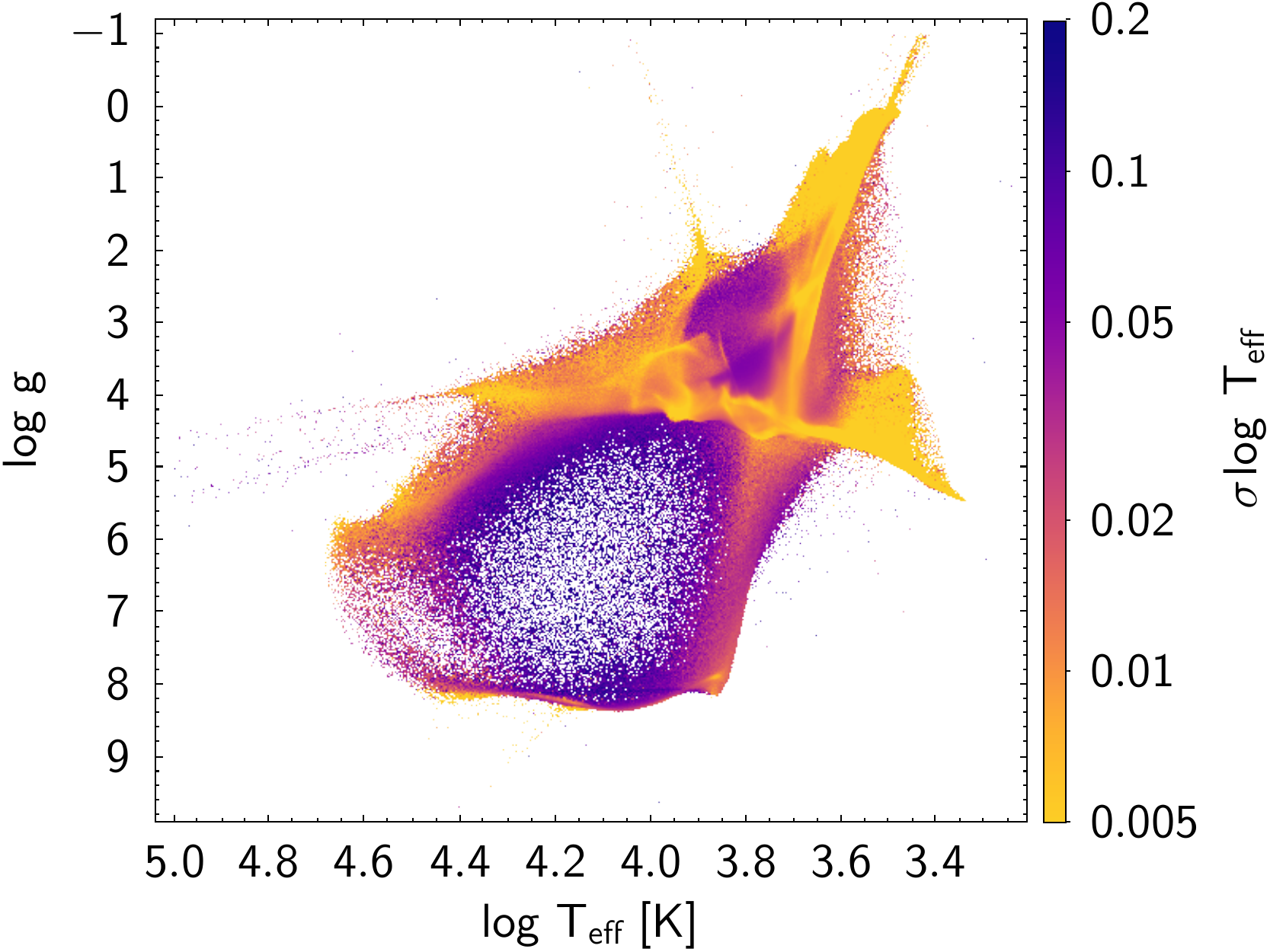}
\plotone{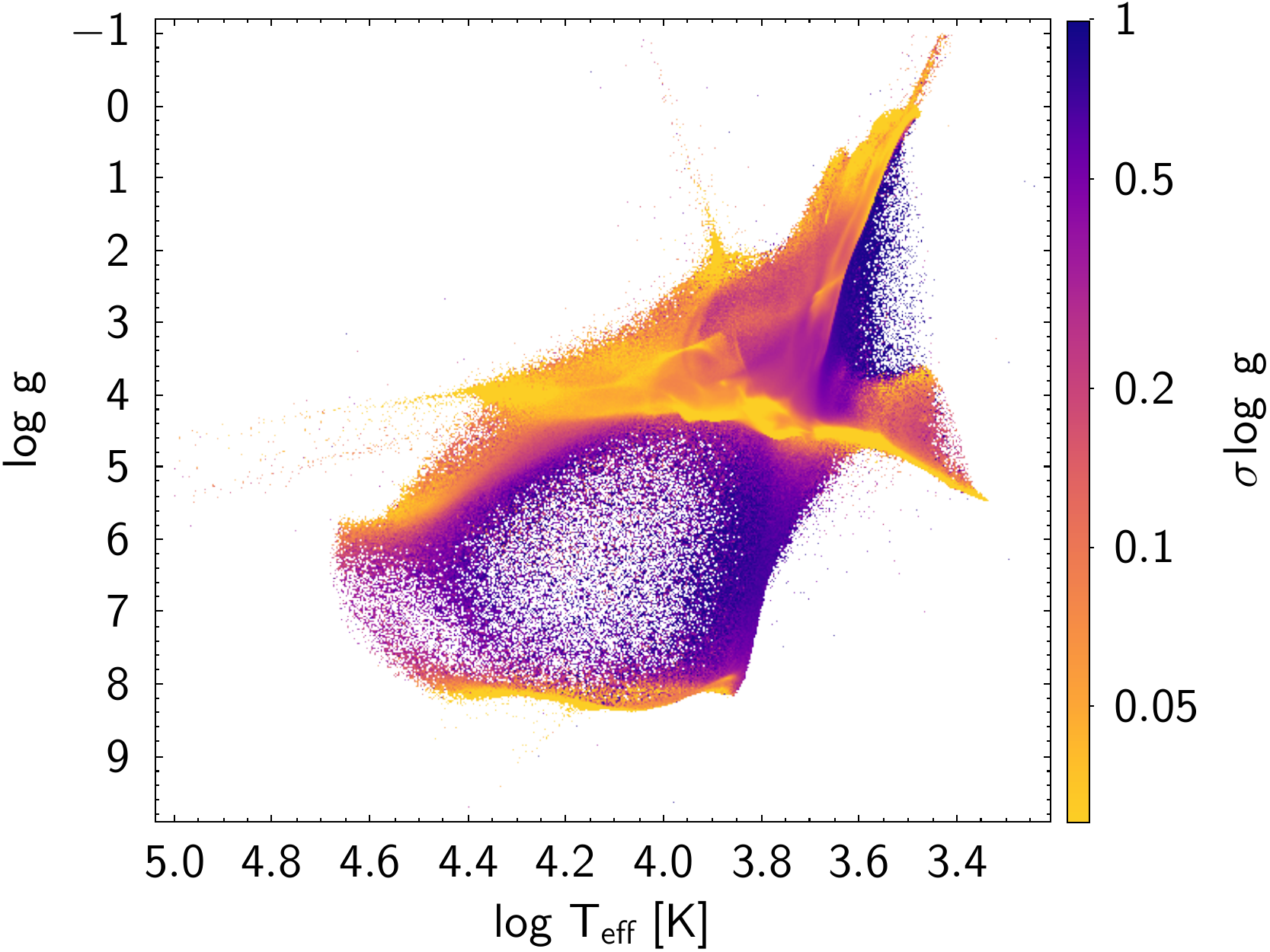}
\plotone{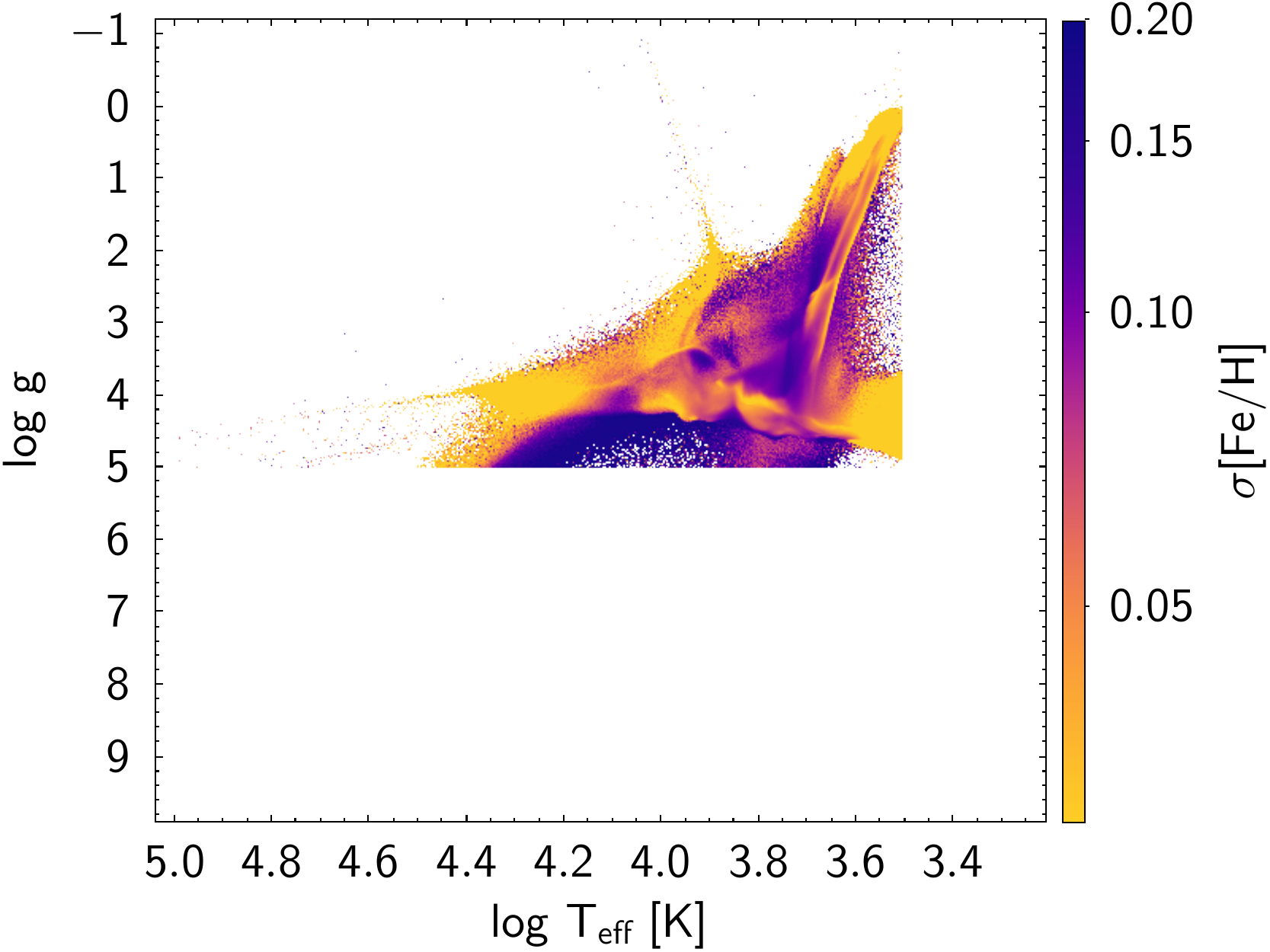}
\caption{Distribution of characteristic uncertainties in the sample.
\label{fig:sigma}}
\end{figure}

\section{Model}
GAIA Net is a 1D residual convolutional neural network based on the architecture of BOSS Net \citep{sizemore2024}. The GAIA Net model takes as input the concatenation of the blue and red photometry coefficients. Because convolutional networks are translation-invariant, we add a fixed positional encoding channel as it provides explicit location information within the photometry coefficients. The photometry coefficients and positional encoding act as inputs to the initial convolutional block, which includes a 1D convolutional layer with a kernel size of 30, batch normalization \citep{batchnorm}, an Exponential Linear Unit (ELU) activation function \citep{ELU}, and a 1D max pooling layer.

The output of the initial convolutional block is fed into the first of three residual blocks. Each residual block contains a 1D convolutional layer with a kernel size of 30, followed by batch normalization, an ELU activation function, a second 1D convolutional layer (also with a kernel size of 30), another batch normalization, a residual connection \citep{he2016deep}, and a final ELU activation function.

Following the three residual blocks, average pooling is applied to the intermediate representation, which is then flattened and passed into a seven-layer fully connected network. Each layer contains 4096 hidden units and is followed by a ReLU activation function. This final fully connected network produces the model predictions: $\log g$, $\log T_{\text{eff}}$, and $[Fe/H]$.

GAIA Net was evaluated using the Mean Squared Error between the true labels and the predictions. The model was trained using the Adamax optimizer \citep{kingma2015} with a learning rate of 0.0004 and a batch size of 512.

\section{Results}

\subsection{Overview}
We have applied Gaia Net to all of the XP spectra that have been made available in Gaia DR3, for a total of $\sim$220 million sources (Figure \ref{fig:predictions}, Table \ref{tab:xp}). Similar efforts for RVS spectra are presented in the Appendix \ref{sec:rvs}.

\begin{deluxetable}{ccl}[!ht]
\tablecaption{Stellar parameters derived from XP spectra
\label{tab:xp}}
\tabletypesize{\scriptsize}
\tablewidth{\linewidth}
\tablehead{
 \colhead{Column} &
 \colhead{Unit} &
 \colhead{Description}
 }
\startdata
source\_id & & Gaia unique identifieer \\
RA & deg & Right ascention in J2000 \\
Dec & deg & Declination in J2000 \\
log \teff & [K] & Effective temperature \\
$\sigma$ log \teff & [K] & uncertainty in log \teff \\
log g &  & Surface gravity \\
$\sigma$ log g & & Uncertainty in log g \\
$\left[\rm{Fe/H}\right]$&  & Metallicity \\
$\sigma$ [Fe/H] & & Uncertainty in [Fe/H] \\
\enddata
\end{deluxetable}

Overall, Gaia Net well represents the underlying shape of the parameter space for all classes of objects. There are some ``unphysical'' features of the parameter space - e.g., the bridge between the cool white dwarfs and the main sequence, such sources were present in the outputs of BOSS Net as well, and usually were indicative of sources with very low signal-to-noise ratio where the model couldn't confidently settle into just a single \logg\ sequence, thus it has split the difference between the two possibilities. Typically these sources have large uncertainties in \logg$>$0.5 dex. While the bulk of these sources are stellar (or substellar) in nature, this population also houses some of the extragalactic sources as well. The bulk of the $\sim$100,000 sources identified as quasars in the Quaia catalog \citep{storey-fisher2024} appear to inhabit this region as well. Given that these sources are incompatible with the model's fundamental assumptions, care should be exercised to exclude these sources when using this catalog, as appropriate. However, given they account to only 0.05\% of the entire sample, the relative degree of contamination is minimal.

There are also a handful of various ``spikes'' protruding away from different corners of the parameter spaces - e.g., extending to very high \teff, or to very low \logg. It is likely that the parameters of these sources may be found just slightly outside of the parameter space covered by the training set, and thus beyond the model's ability to generalize. However, such sources are rare, there are only $\sim$1000 of stars the predicted stellar parameters of which are found outside of the bounds of the Figure \ref{fig:predictions}.

The typical scatter in \logg\ between the labels and the predictions is 0.135 dex, increasing to $\sim$0.2 in the most challenging corners of the parameter space (e.g., the base of the red giant branch, or for high mass stars). Typical scatter in \logteff\ is 0.01 dex for low mass stars, and 0.025 dex for high mass stars. For [Fe/H] it is 0.13 for low mass stars and 0.22 dex for high mass stars (Figure \ref{fig:errors}). This subset does not exhibit the same artificial ``bridge'' between the main sequence and the white dwarf sequence as the full sample, because it is restricted solely to the sources for which originally there were BOSS Net labels, i.e., predominantly higher signal-to-noise sample.

The stellar parameters returned by the model are deterministic, i.e., the same input data would always produce the same prediction. This makes it difficult to evaluate the uncertainties. In BOSS Net we have estimated the errors through passing different representations of the same spectra scattered by the flux errors to the model multiple times. Similar approach has been applied in here, taking into the account the covariance matrix. Figure \ref{fig:sigma} shows the distribution of uncertainties in the sample.

\subsection{Comparison to other pipelines}

\begin{figure*}
\epsscale{1.0}
		\gridline{\leftfig{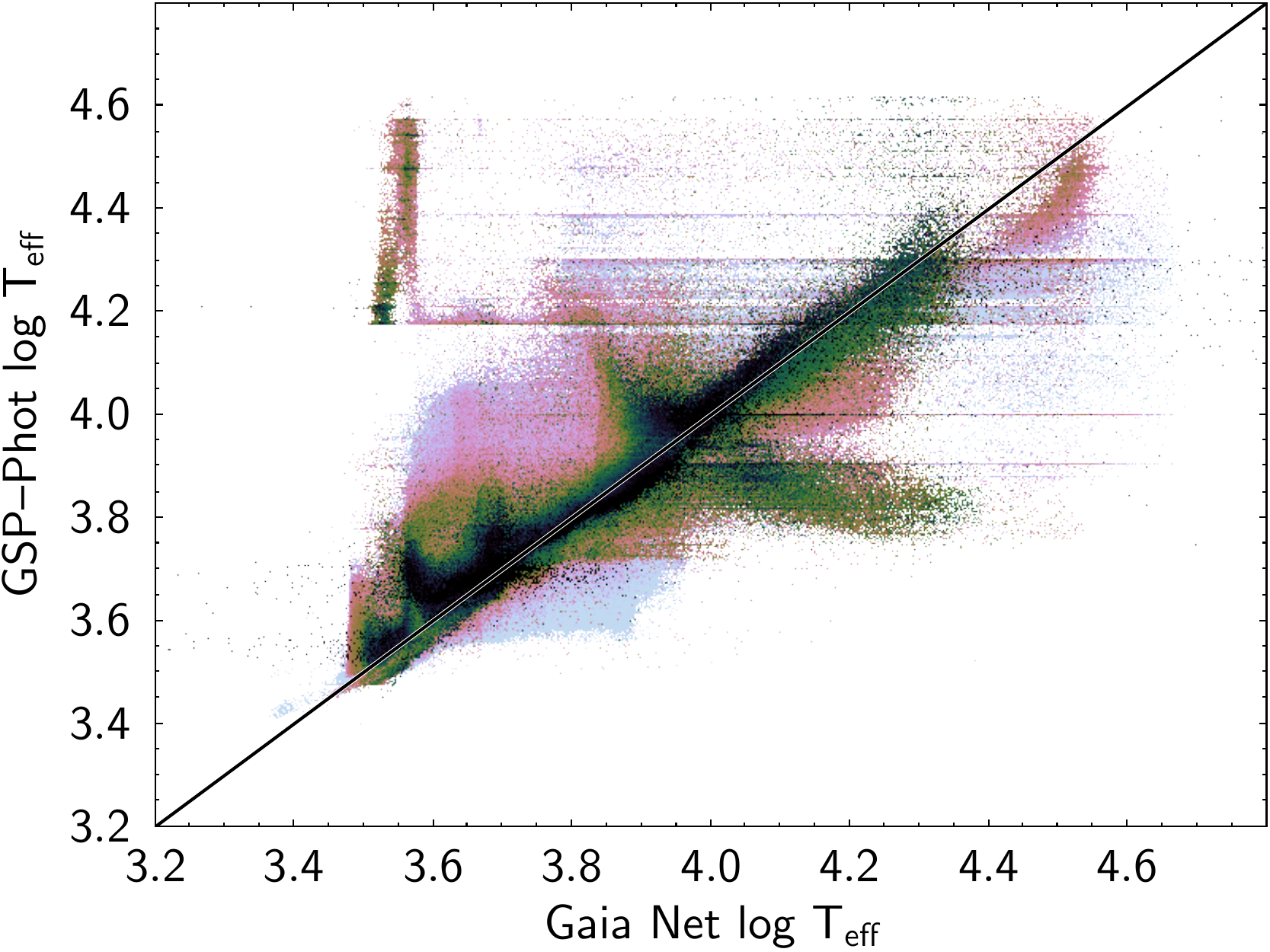}{0.33\textwidth}{}
                 \leftfig{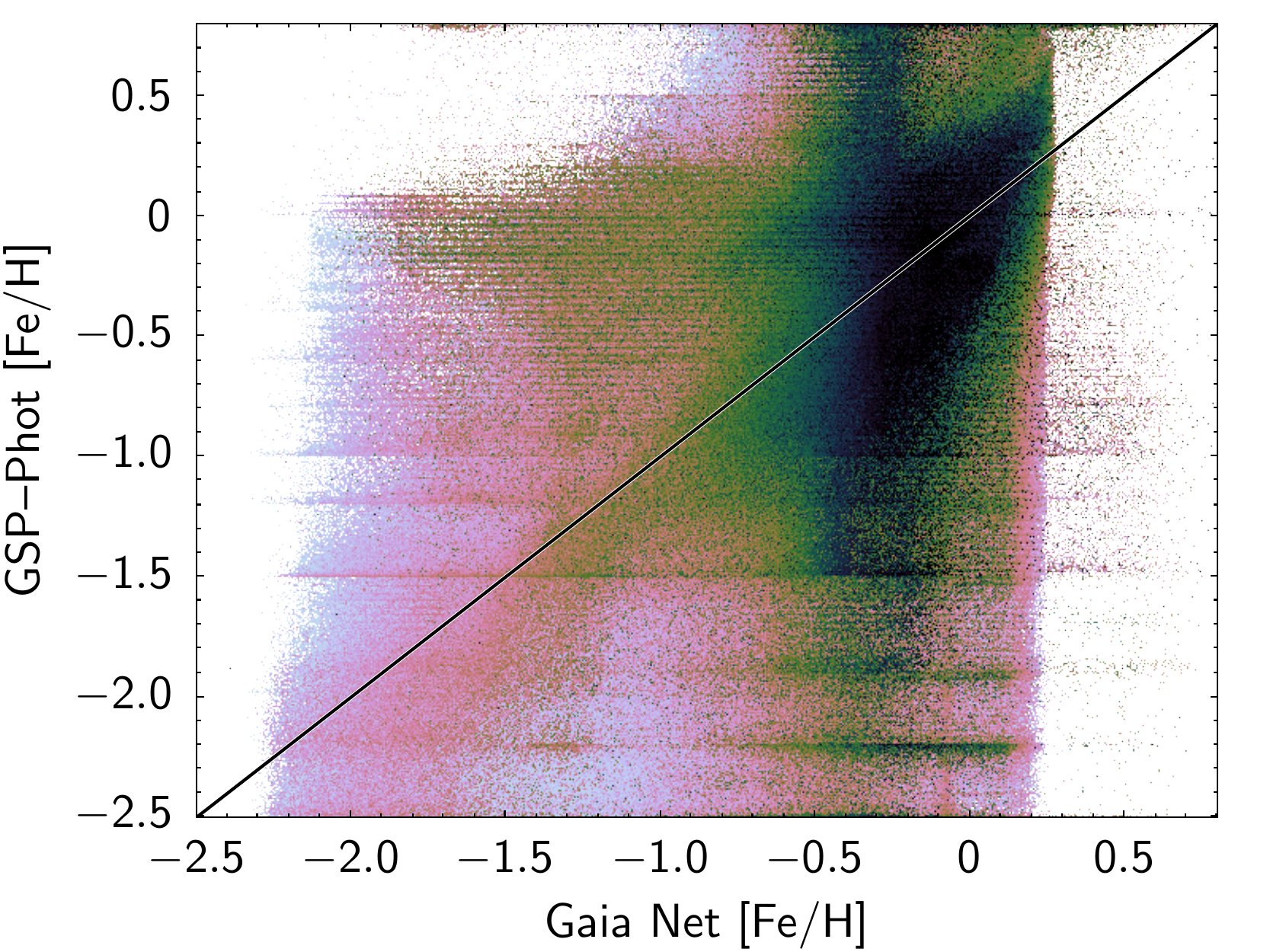}{0.33\textwidth}{}
		          \leftfig{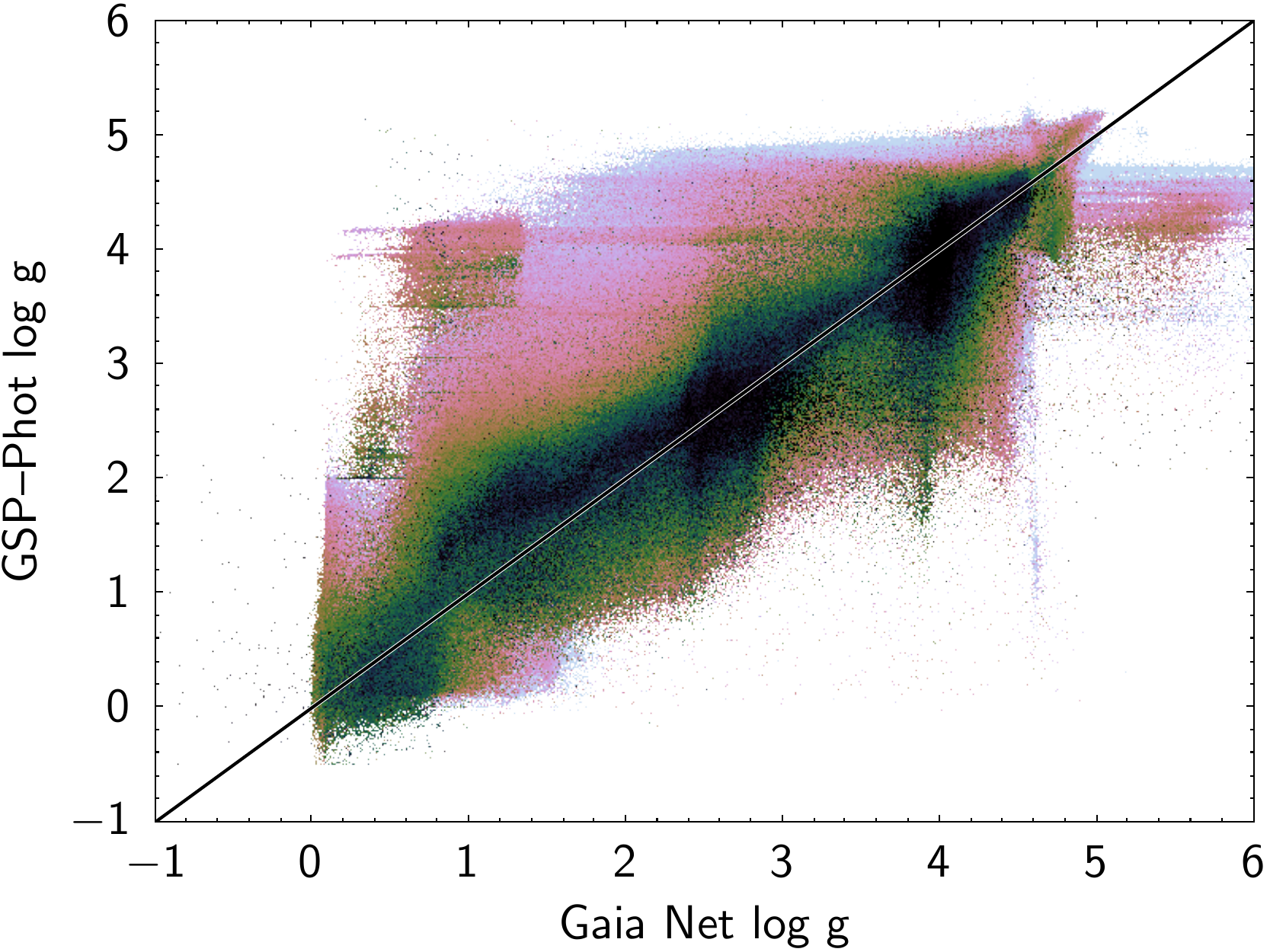}{0.33\textwidth}{}
        }
        \vspace{-0.8cm}
        \gridline{\leftfig{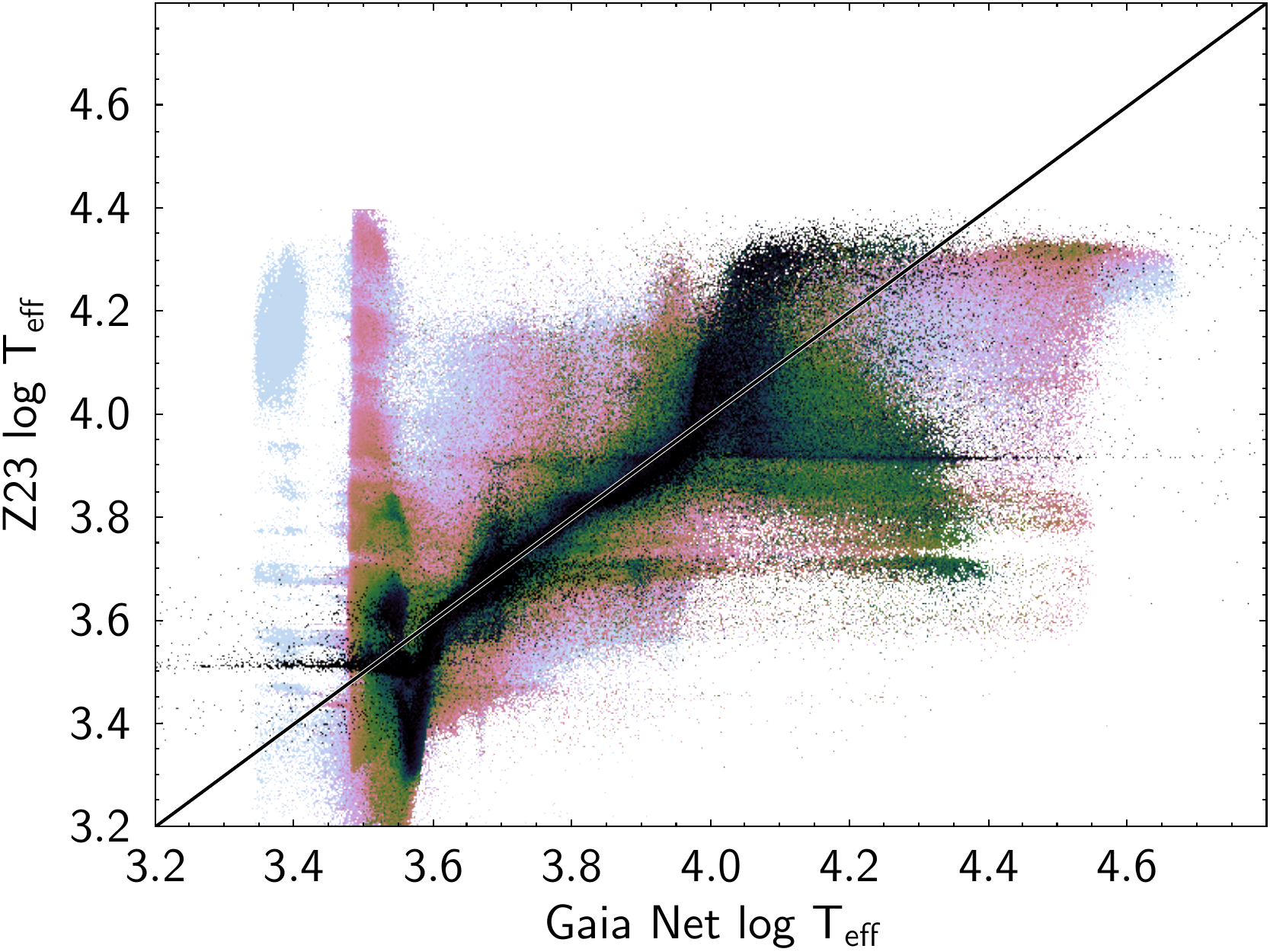}{0.33\textwidth}{}
		          \leftfig{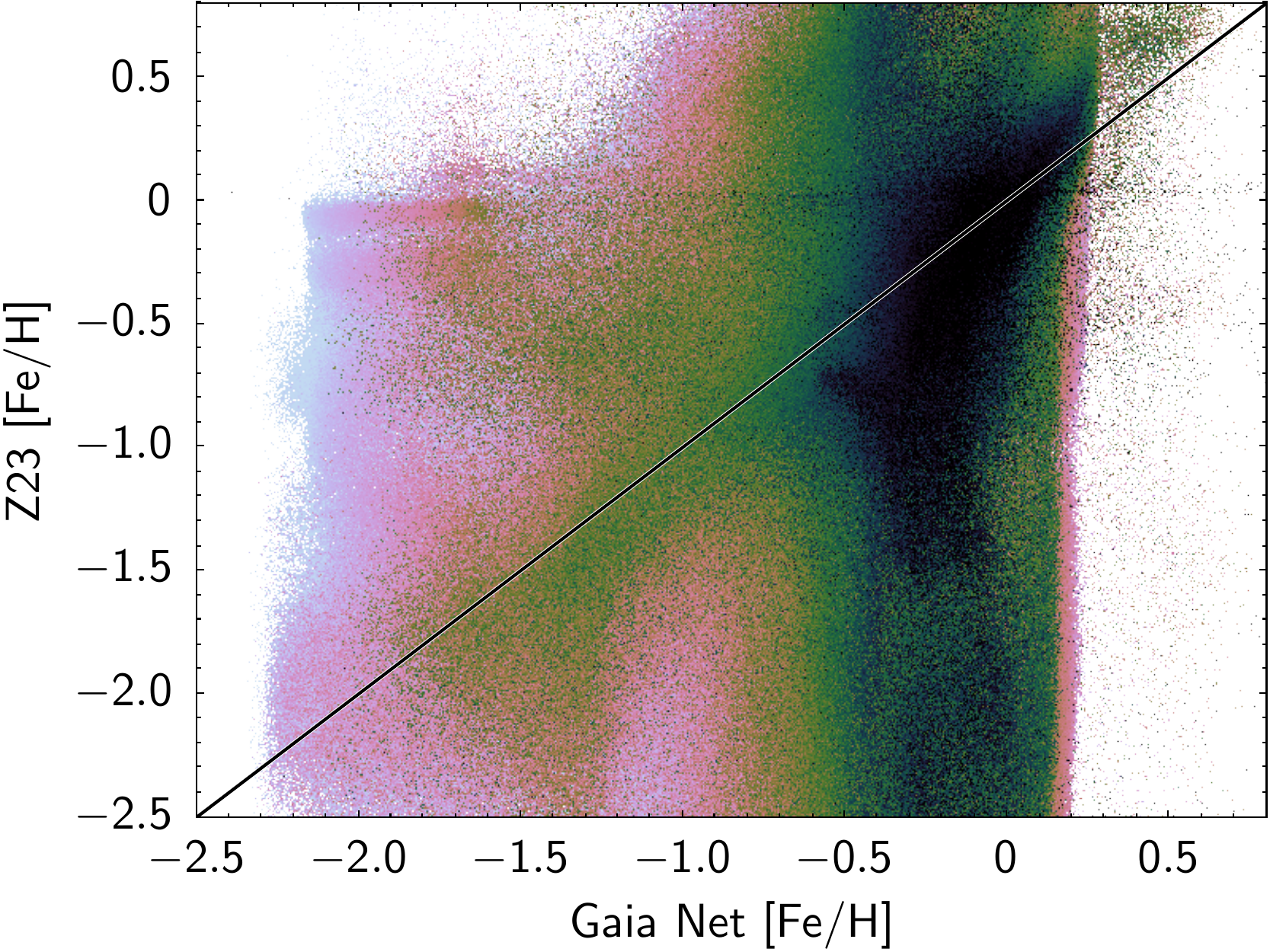}{0.33\textwidth}{}
                \leftfig{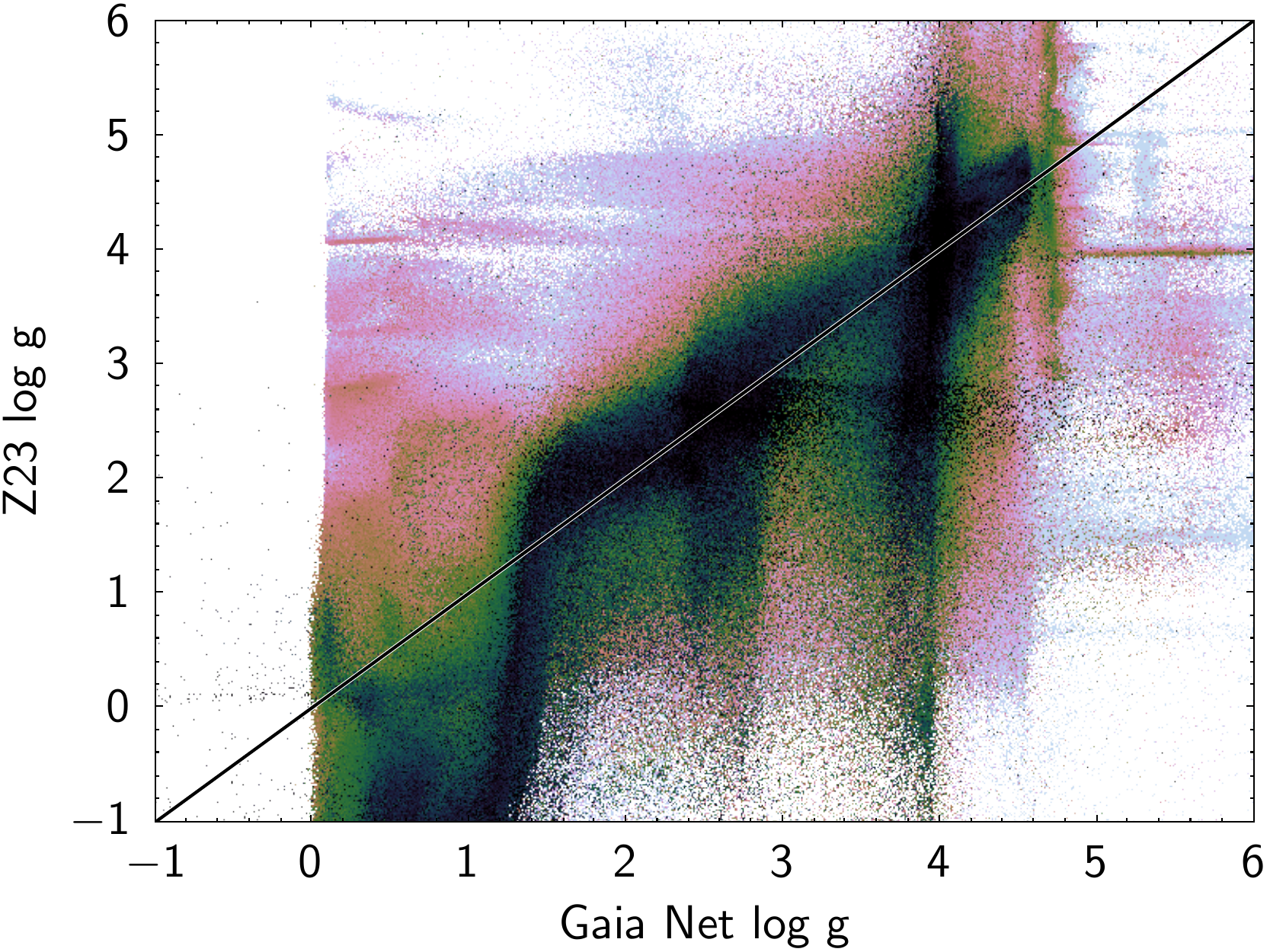}{0.33\textwidth}{}
        }\vspace{-0.8cm}
        \gridline{\leftfig{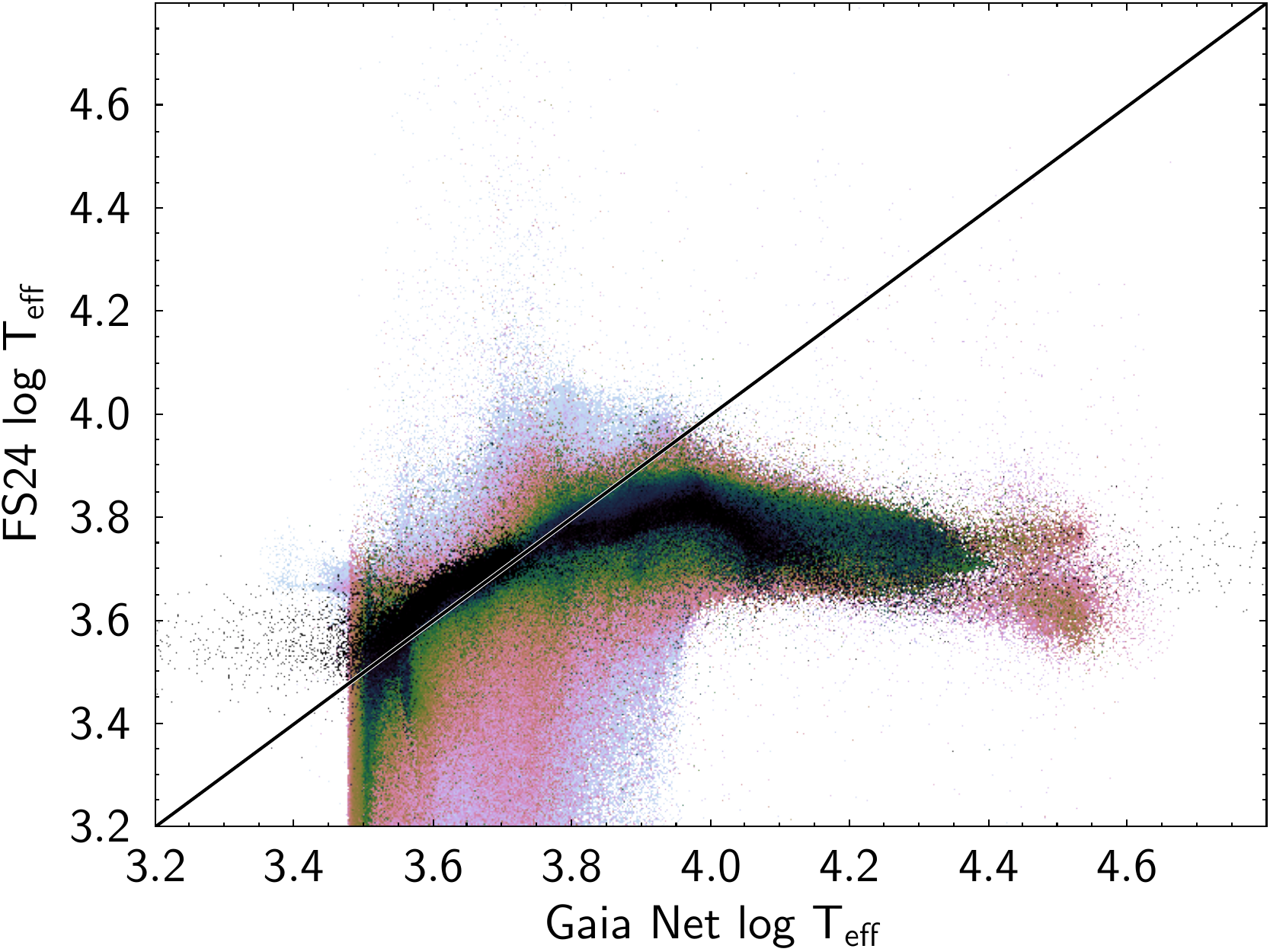}{0.33\textwidth}{}
		          \leftfig{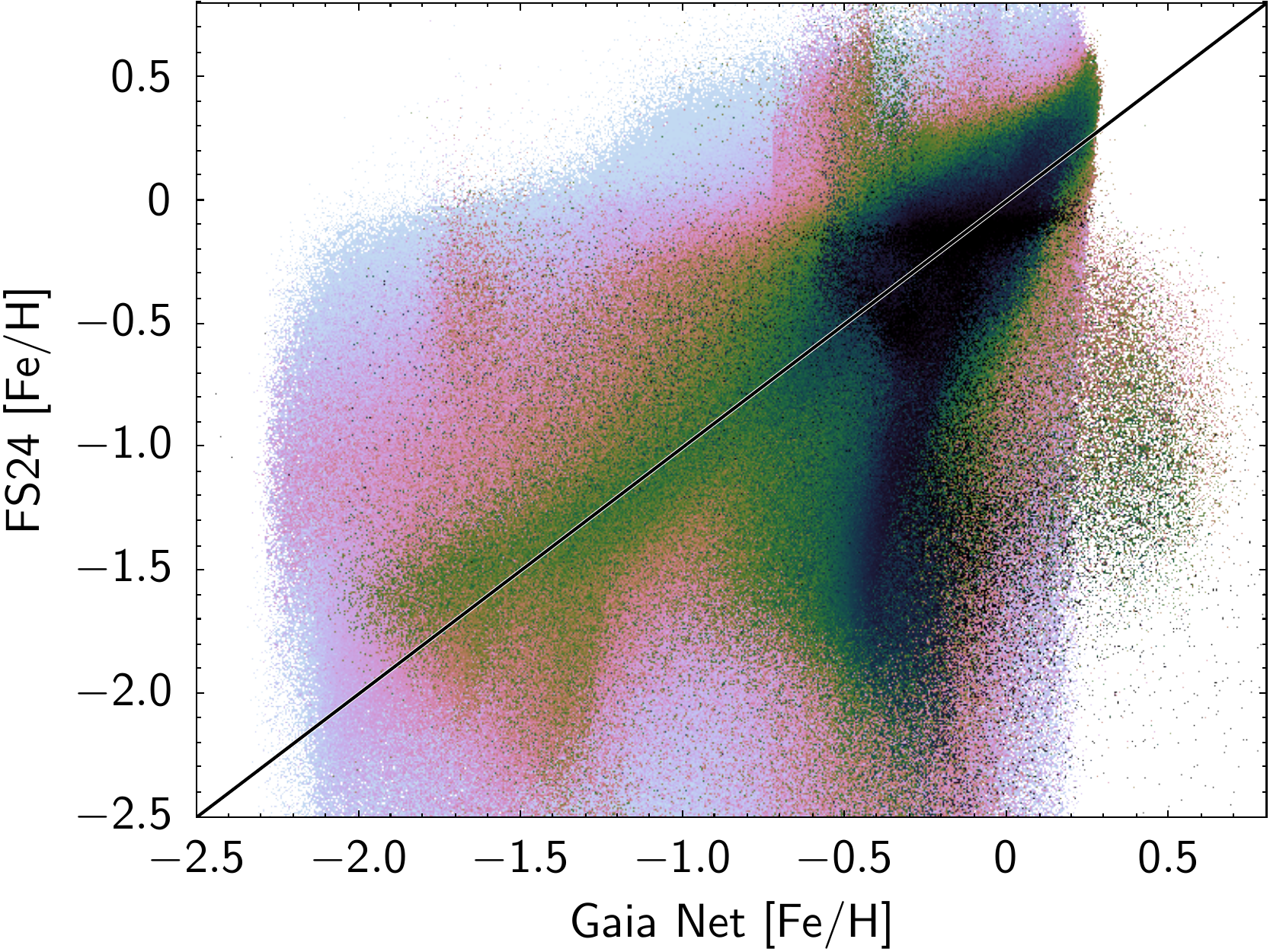}{0.33\textwidth}{}
                \leftfig{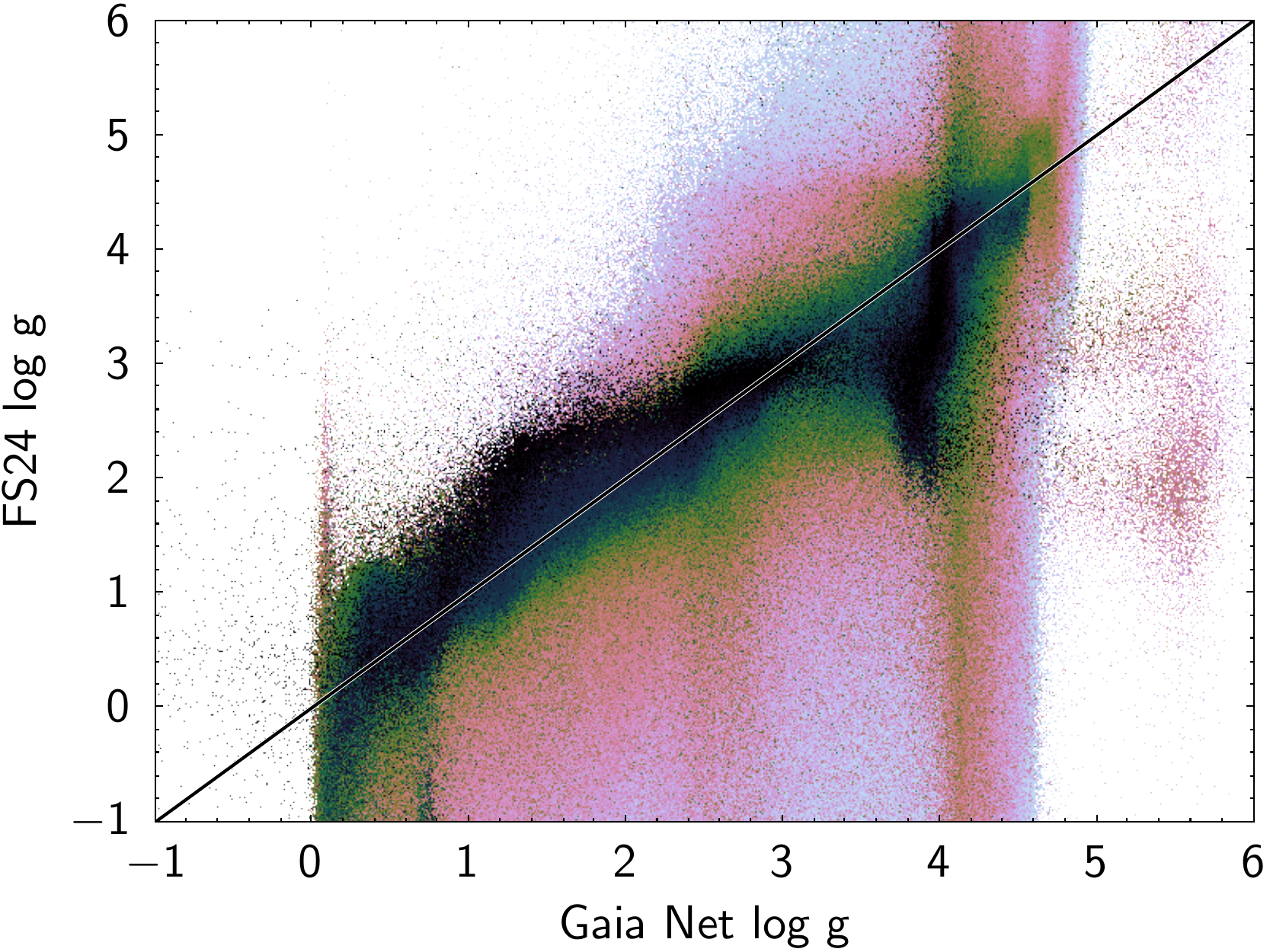}{0.33\textwidth}{}
        }\vspace{-0.8cm}
        \gridline{\leftfig{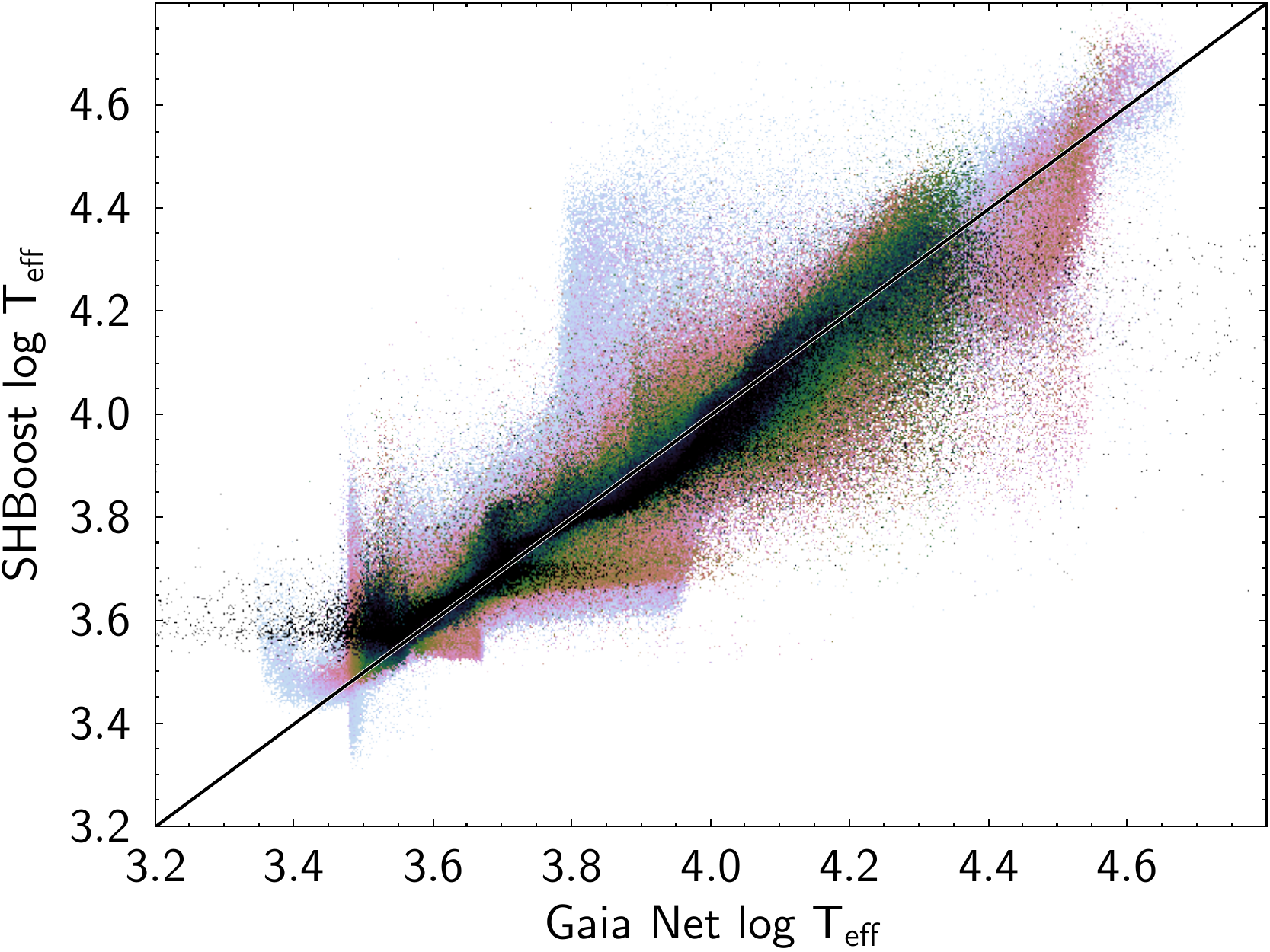}{0.33\textwidth}{}
		          \leftfig{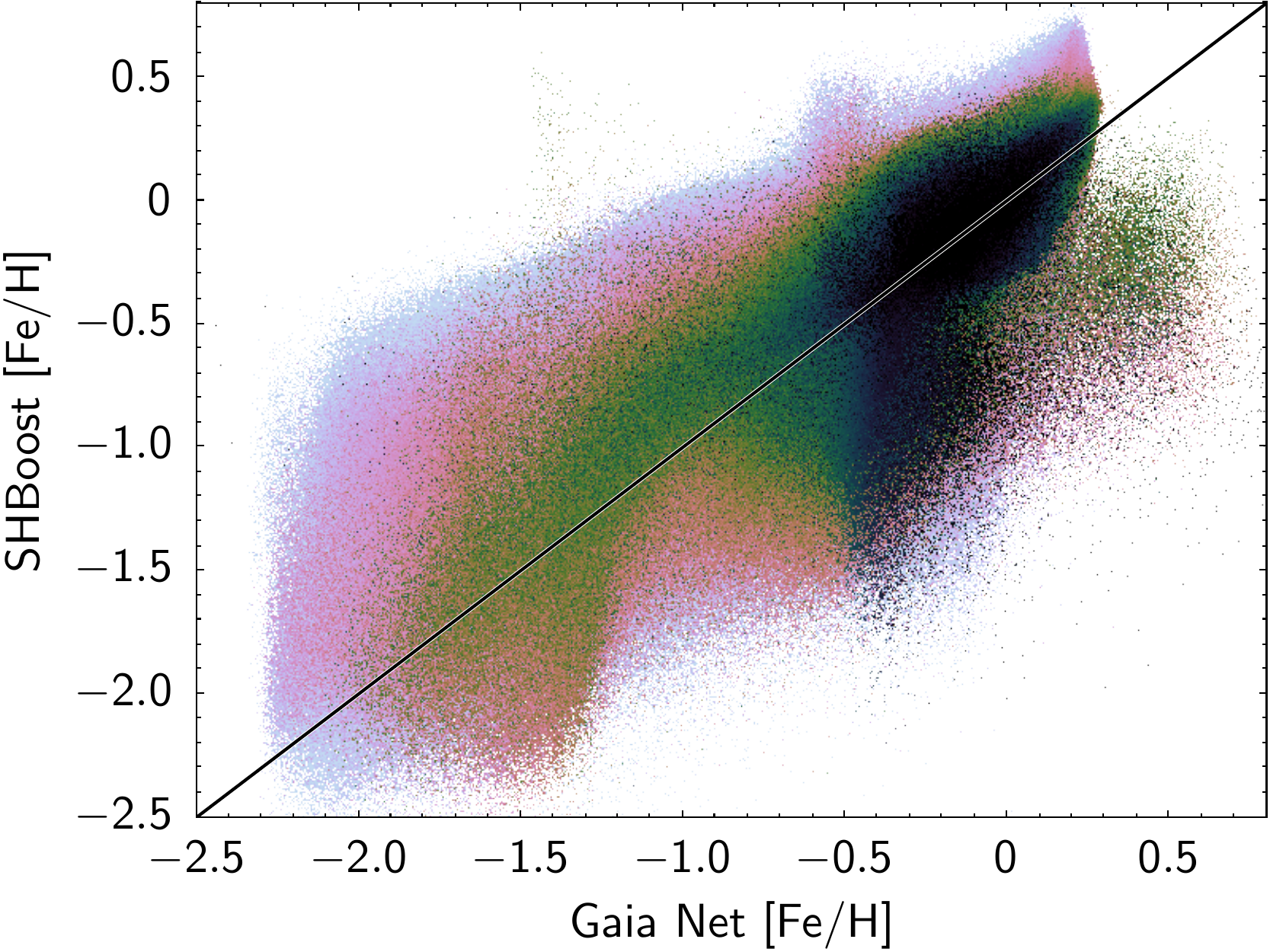}{0.33\textwidth}{}
                \leftfig{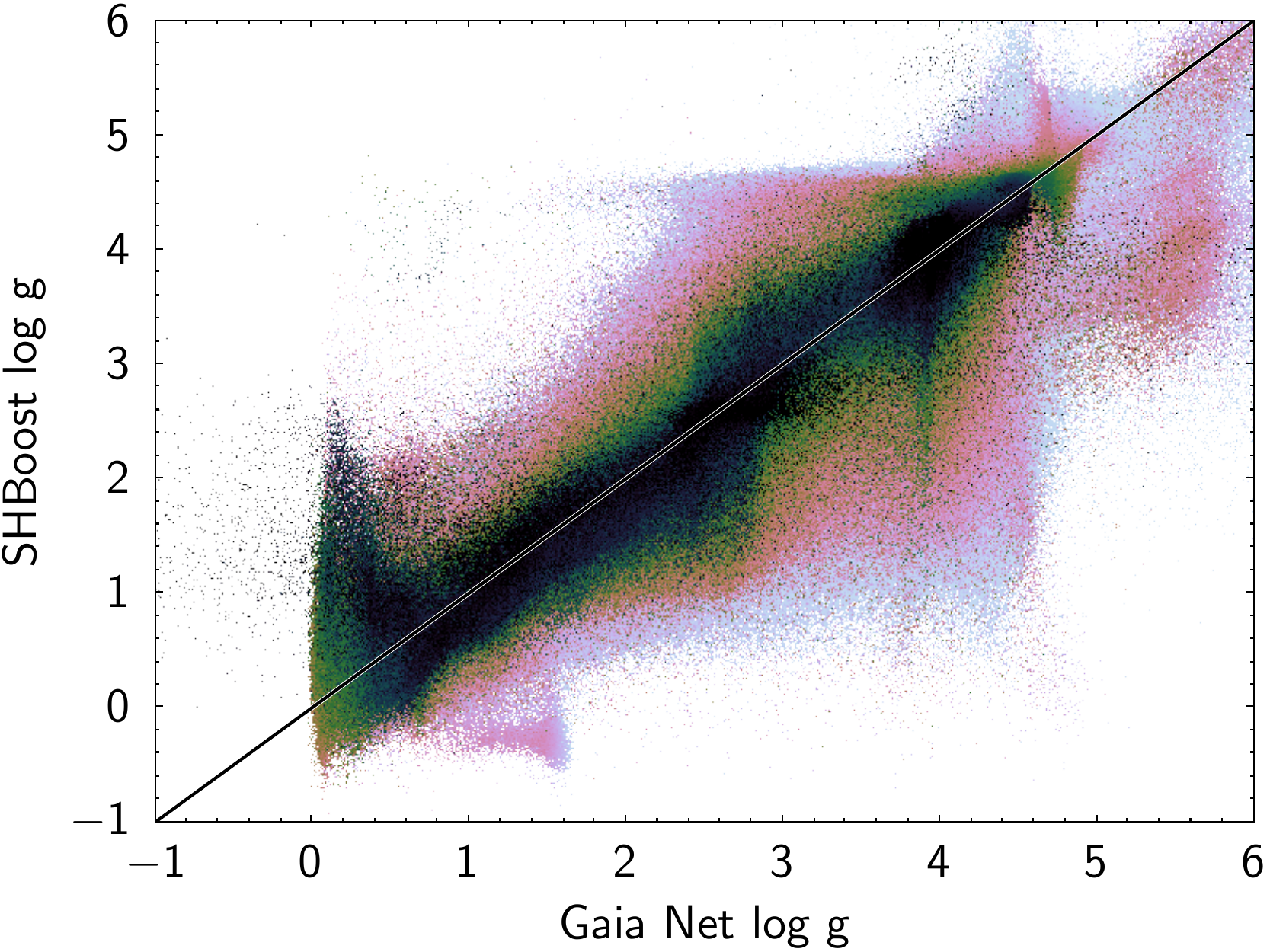}{0.33\textwidth}{}
        }\vspace{-0.8cm}
        \plotone{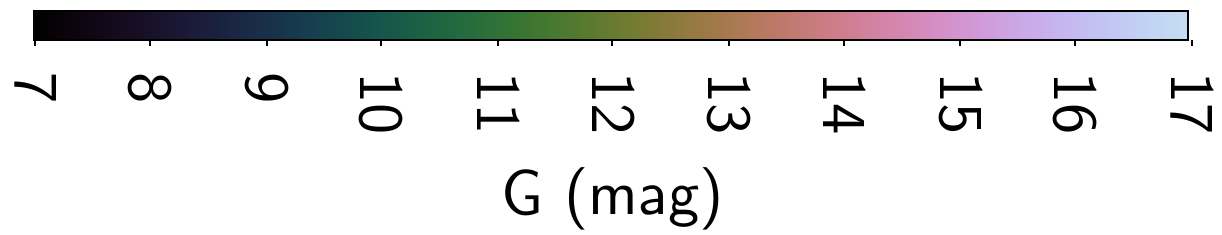}
\caption{Comparison of the stellar parameters extracted Gaia Net and other works. First row: Gaia DR3 GSP-Phot pipeline \citep{andrae2023}. Second row, \citet{zhang2023}. Third row, \citet{fallows2024}. Fourth row, SHBoost pipeline \citep{khalatyan2024}.
\label{fig:comparison}}
\end{figure*}

We compare Gaia Net to the outputs from other works that have processed XP spectra (Figure \ref{fig:comparison}), such as the ``native'' GSP-Phot stellar parameters that have been released as part of Gaia DR3 \citep{andrae2023}, as well as to the subsequent independent works from \citet{zhang2023}, \citet{fallows2024}, and \citet{khalatyan2024}.  

For the most part, they all have a comparable performance for very bright sources, with $G<10$ mag over the valid range of parameters. The catalog from \citet{fallows2024} did not have the training sample extend to early type stars and thus underestimates their \teff. Similarly, both \citet{zhang2023} and GSP-Phot appear to have strong systematic artifacts near the edges of the \teff\ grid. \citet{zhang2023} also appear to lack reliable \logg\ predictions for AGB stars with \logg$<$1. 

And, although all catalogs appear to have comparable [Fe/H] for late type stars (at least for the bright sources), they all appear to favor lower [Fe/H] of early type stars in comparison to Gaia Net. Due to a sparsity of metal-sensitive features in hot stars, let alone in spectra with such low resolution, measuring reliable [Fe/H] in these sources can be a challenge, and all catalogs demonstrate a difference in [Fe/H] distribution between hot and cool stars. However, in Gaia Net, the difference between the typical [Fe/H] between stars with \teff$>10,000$ K and \teff$<10,000$ K is only 0.3 dex, whereas in SHBoost \citep{khalatyan2024}, this difference is 0.6 dex, and in \citet{fallows2024} it is 1.1 dex.

SHBoost appears to have the most similar performance to Gaia Net, both in the variety of different source types evaluated here, as well as the comparability of the predictions. The scatter in the predicted parameters increases for fainter sources in all pipelines, but even in the fainter stars, but despite this scatter, predictions from SHBoost appear to be strongly correlated with the predictions from Gaia Net, even for sources with $G>17$ mag.

While overall the performance is comparable, there are some source types for which either pipeline offers more robust measurements. E.g., comparing predictions of \teff\ \& \logg\ against a color-magnitude diagram, SHBoost appears to make more robust measurements of \logg\ on very cool white dwarfs (likely in part because of the inclusion of the integrated photometry and astrometry), whereas Gaia Net ends up confusing their placement between the main sequence and the white dwarf cooling sequence.

Gaia Net may have somewhat peculiar parameters on very metal poor OB stars, i.e., it appears to shift 
 \logg\ values of all of the hottest stars in the Magellanic Clouds (MCs) to coincide with the location of the subdwarfs. This does not appear to affect hot stars in the Milky Way (MW), well separating the massive stars and the hot subdwarfs (which can be thought of as extreme horizontal branch stars). On the other hand, SHBoost produces more comparable parameters for hot stars both in MCs and in MW, although the resulting \logg\ for almost a half of the OB stars in MCs is also found below the main sequence.

On the other hand, Gaia Net significantly improves over SHBoost with regards to the low mass stars. Stars cooler than 2800 K appear to be absent in the training set for SHBoost, as such it significantly overestimates their \teff, anti-correlating it against the trend in colors. In contrast, Gaia Net can go down in \teff\ to $\sim$2000 K, too the coolest sources for which XP spectra are available.

SHBoost also cannot discriminate parameters of the pre-main sequence stars, confusing their \logg\ for the main sequence sources. This is generally the case for all of the pipelines and models that have processed XP spectra, outside of Gaia Net, although \citet{fallows2024} do have a slight sensitivity towards them.

\subsection{Test of performance of pre-main sequence stars}

\begin{figure*}
\epsscale{1.15}
        \plottwo{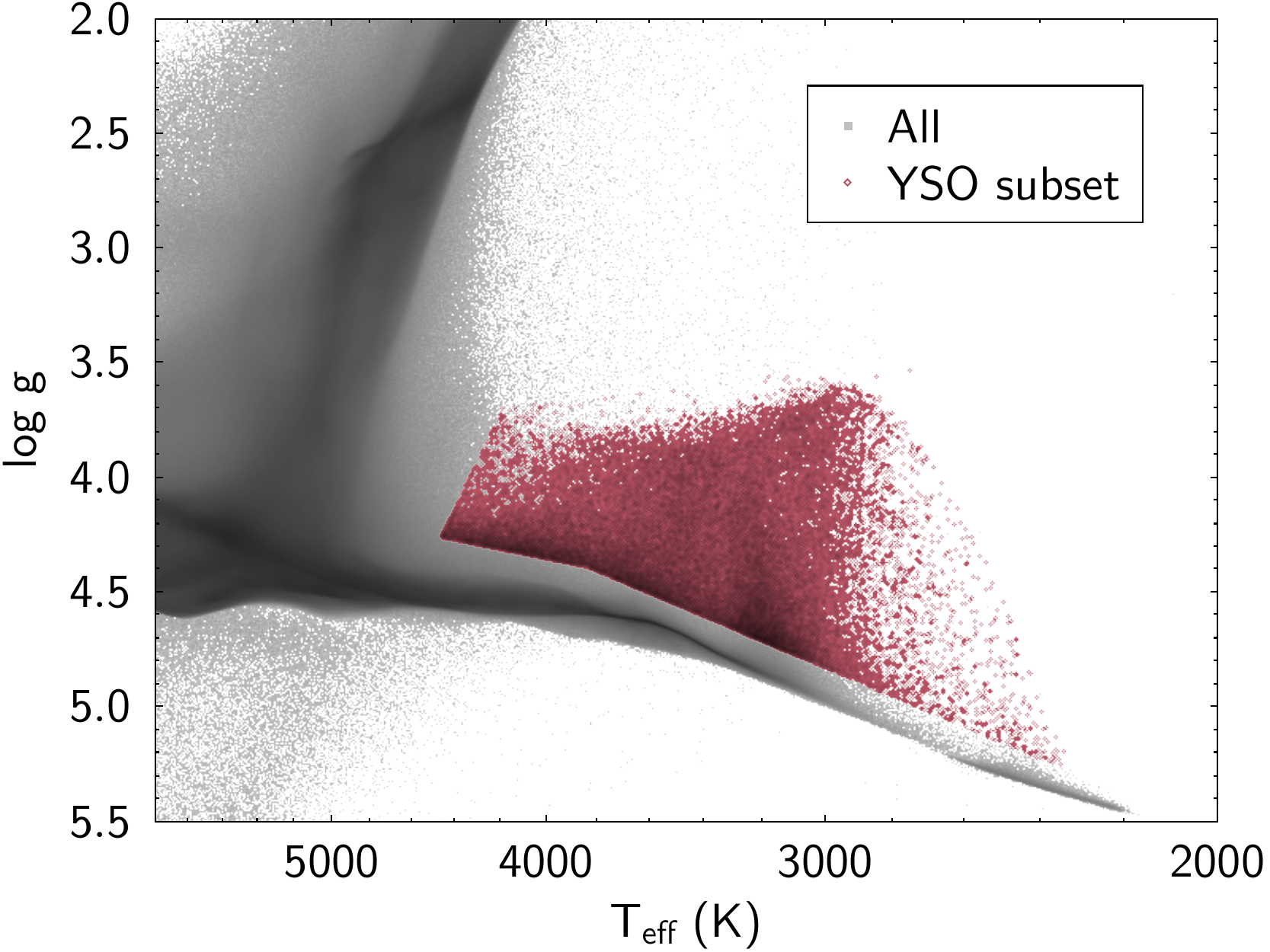}{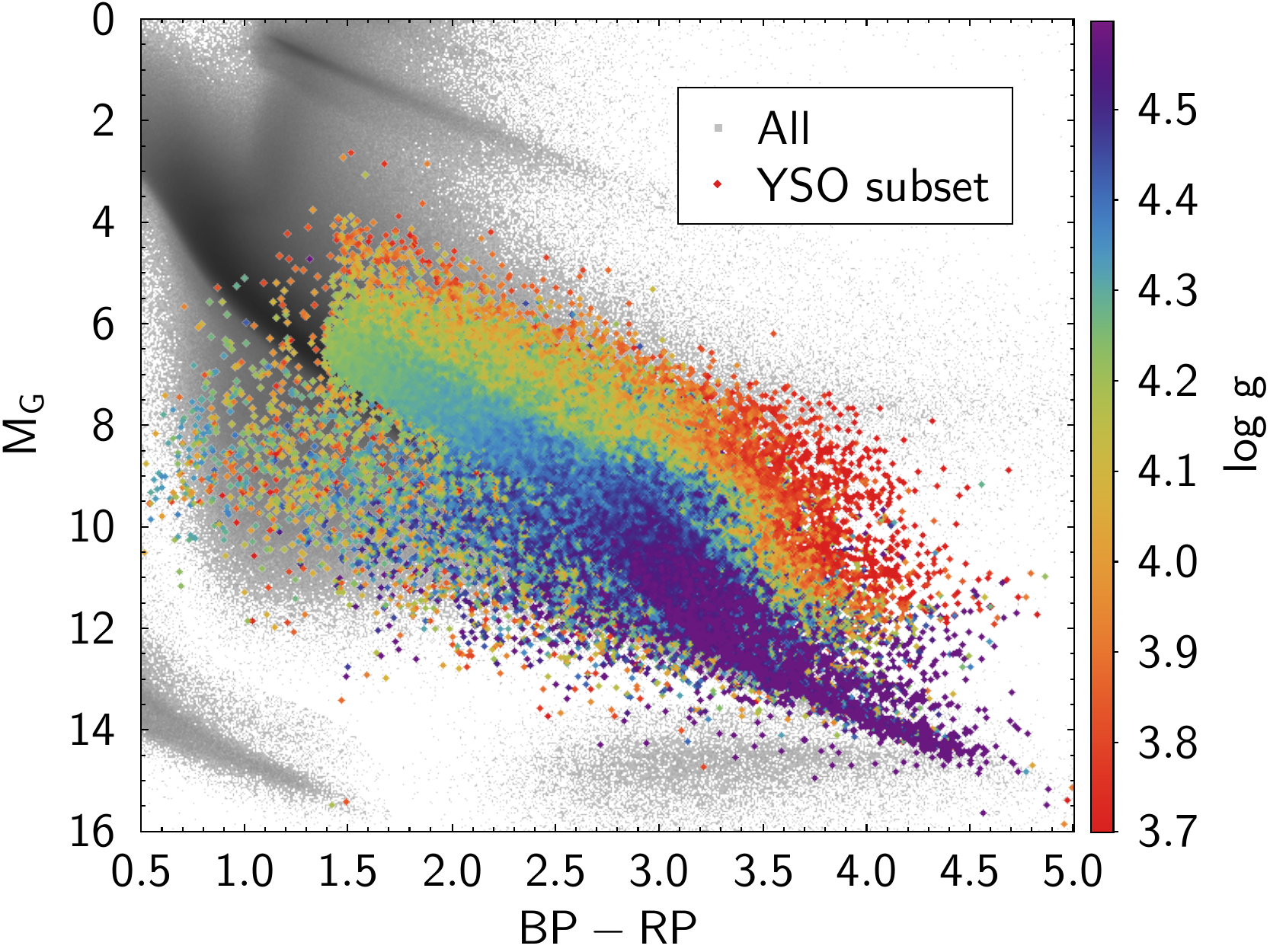}
        \plotone{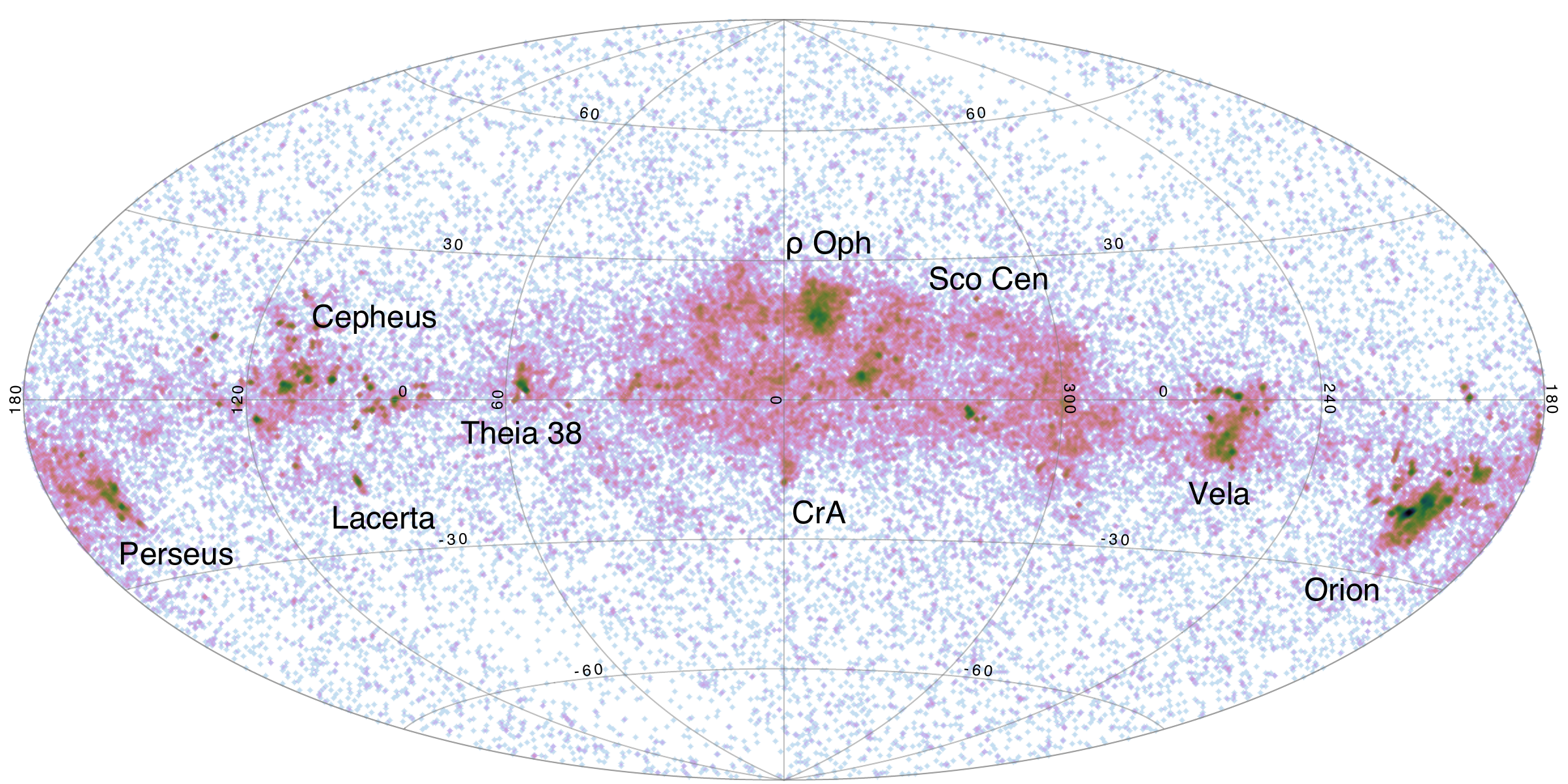}
        \plottwo{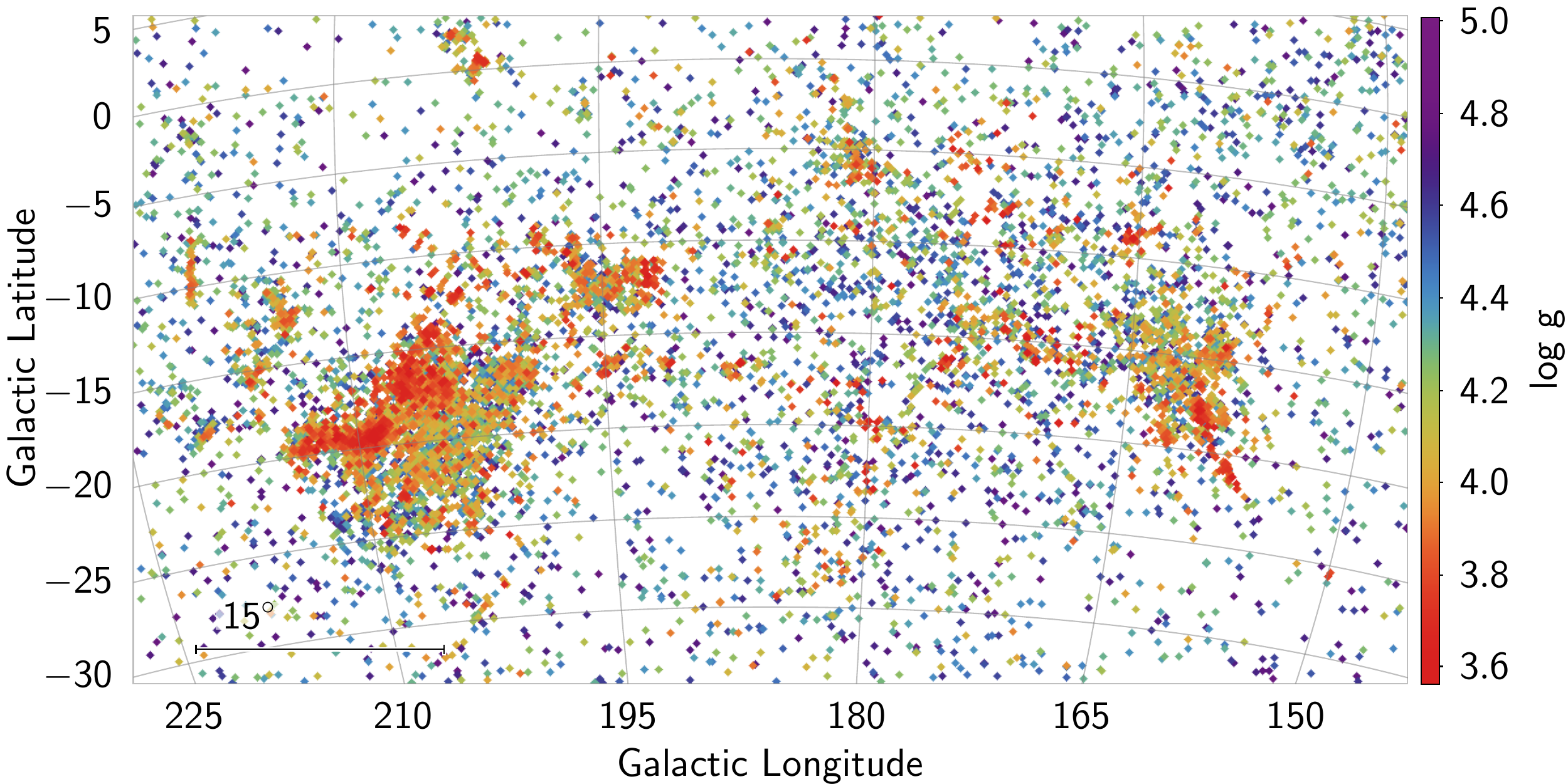}{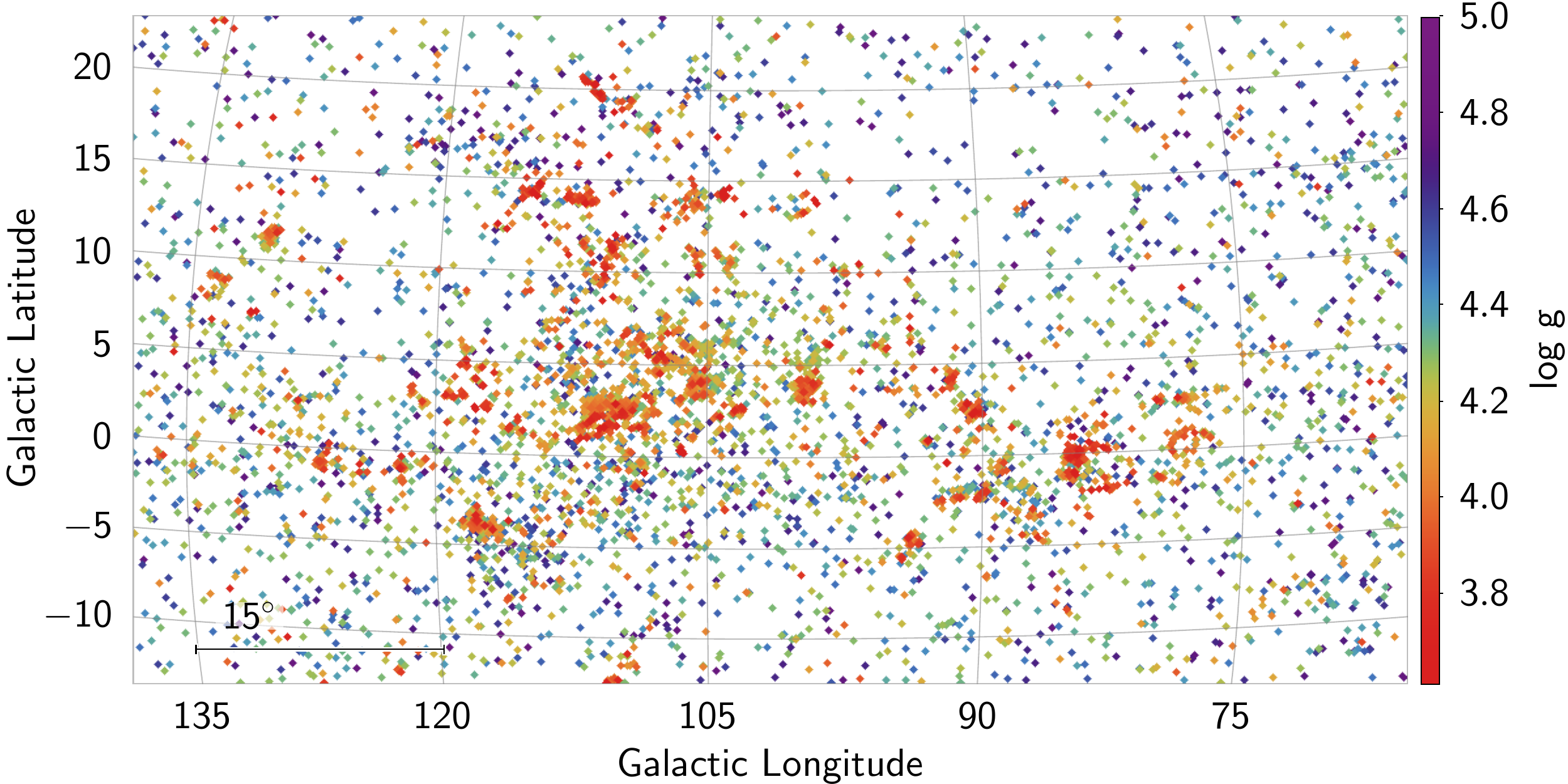}
\caption{Selection of pre-main sequence stars based on spectroscopic parameters produced by Gaia Net. Top left: \teff\ \& \logg\ selection. Top right: HR diagram produced by this selection, color-coded by \logg. Middle: Sky projection of the selected pre-main sequence candidates, color-coded by their density. Bottom: zoom in sky projection towards Orion \& Perseus (left), as well as Cepheus clouds (right). Sources are color coded by \logg\ as a proxy of age (sources with low log g have larger radii and they are less evolved).
\label{fig:ysos}}
\end{figure*}

Often times, low mass young stars are identified and characterized on the basis of their photometry. Being larger than main sequence stars and rapidly contracting, they begin their lives as highly luminous, incrementally decreasing in their luminosity as they get older. Through comparing the position of young stars on a color-magnitude diagram, it is possible to determine their ages \citep[e.g.,][]{mcbride2021}. However, there are challenges associated with this. Colors are often affected by extinction, making it not only difficult to firmly place a given source on a particular isochrone, but also there is often contamination from the reddened field stars. On top of that, there is an issue of the binary sequence. Binaries with equal masses are more luminous than single stars. Unlike in the evolved clusters, however, where the binary sequence is easily apparent, in star forming region with a meaningful age spread of even a few Myr, the evolution of luminosity is too rapid, instead their ages are systematically underestimated.

Spectroscopy offers a significant advantage on both of these fronts. \teff\ allows to study sources independent of their extinction. And, given that young stars contract over time, their ages are strongly correlated with \logg. Unlike luminosity, however, \logg\ measurements are not significantly biased by binary stars: equal mass binaries produce the same spectroscopic template. At resolution offered by Gaia XP spectra, there are no double-lined spectroscopic binaries that would meaningfully affect the parameter determination. And, upon examining sources with large Renormalised Unit Weight Error (RUWE), or known binary stars, we see no systematic difference in their \teff\ and \logg\ compared to the other stars.

Thus, to demonstrate the performance of Gaia Net on the young stars, we select a parameter space of sources cooler than 4500 K that are generally located above the main sequence, avoiding the red giant branch (Figure \ref{fig:ysos}). This selection was further restricted to sources sources with parallax $>1$ and fidelity\_v2$>0.5$ flag from \citet{rybizki2022} . Although it did exclude some of the sources found in Perseus and Sagittarius spiral arms, this restriction did not have a significant impact on the selection of the sources in the Solar Neighborhood, and it minimized the contamination directly towards the Galactic center from the red giants in the confusion limited regime where, due to difficulty of keeping track of multiple sources in such close proximity, they were erroneously reported to be nearby. Some of these sources might scatter into the selected \teff\ \& \logg\ parameter space, especially if their spectra are a blend of several stars.

We examine the overall distribution of the selected sources in the plane of the sky, and the resulting structure seems to cleanly trace the distribution of the nearby star-forming regions, with populations associated with Sco-Cen, Vela, Orion, Perseus, and Cepheus, and many others, being easily apparent. Indeed, in case of Cepheus, instead of being a monolithic star forming region it is a collection of several small clouds that are distinct spatially and kinematically \citep{szilagyi2023}, and even a rough selection in \teff\ and \logg\ makes all of the individual clusters readily apparent. We note that all of these populations do appear to have expected parallax distribution, and there does not appear to be any obvious contamination from stars significantly behind the population, including in the regions of very high extinction such as $\rho$ Oph or Orion A. This is in contrast to a purely photometric selection \citep[e.g.,][]{mcbride2021}.

The ability to discriminate YSOs in this sample appears to extend to $\sim$20--30 Myr in this sample, as in older stars \logg\ inferred using Gaia Net begin to blend in with the main sequence. In higher resolution spectra it is possible to push this limit to 40--50 Myr \citep{sizemore2024,saad2024}. However, it is already quite remarkable to have robust \logg\ measurements obtained directly from XP spectra, even up to a lower age limit, without any additional metadata.

These \logg s do not only make it possible to identify YSOs, they also have sensitivity to the underlying age of a young star. E.g., in Orion and Perseus, regions that are known to be younger, such as Orion A \& B molecular clouds \citep{kounkel2018a}, as well as NGC 1333 and IC 345 \citep{kounkel2022}, do indeed appear to have lower \logg\ than the rest of the parent population. A similar trend persists in the other star forming regions. And, examining the selection on the HR diagram photometrically, color-coding all of the sources by their measured \logg, it is possible to see a gradient, from the most luminous stars having lower \logg, as is expected through pre-main sequence evolution.

\begin{figure}
\epsscale{1.25}
        \plotone{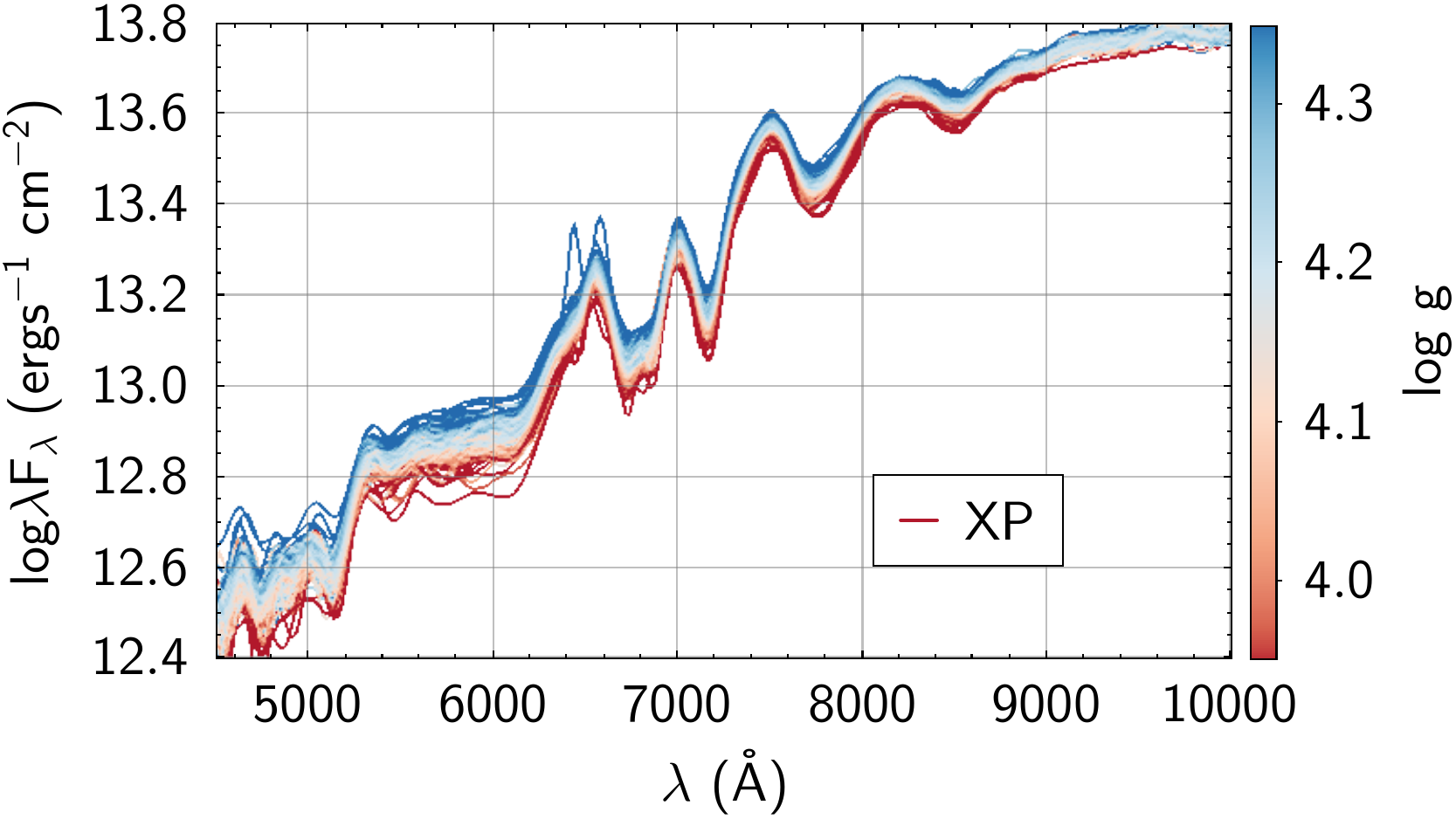}
        \plotone{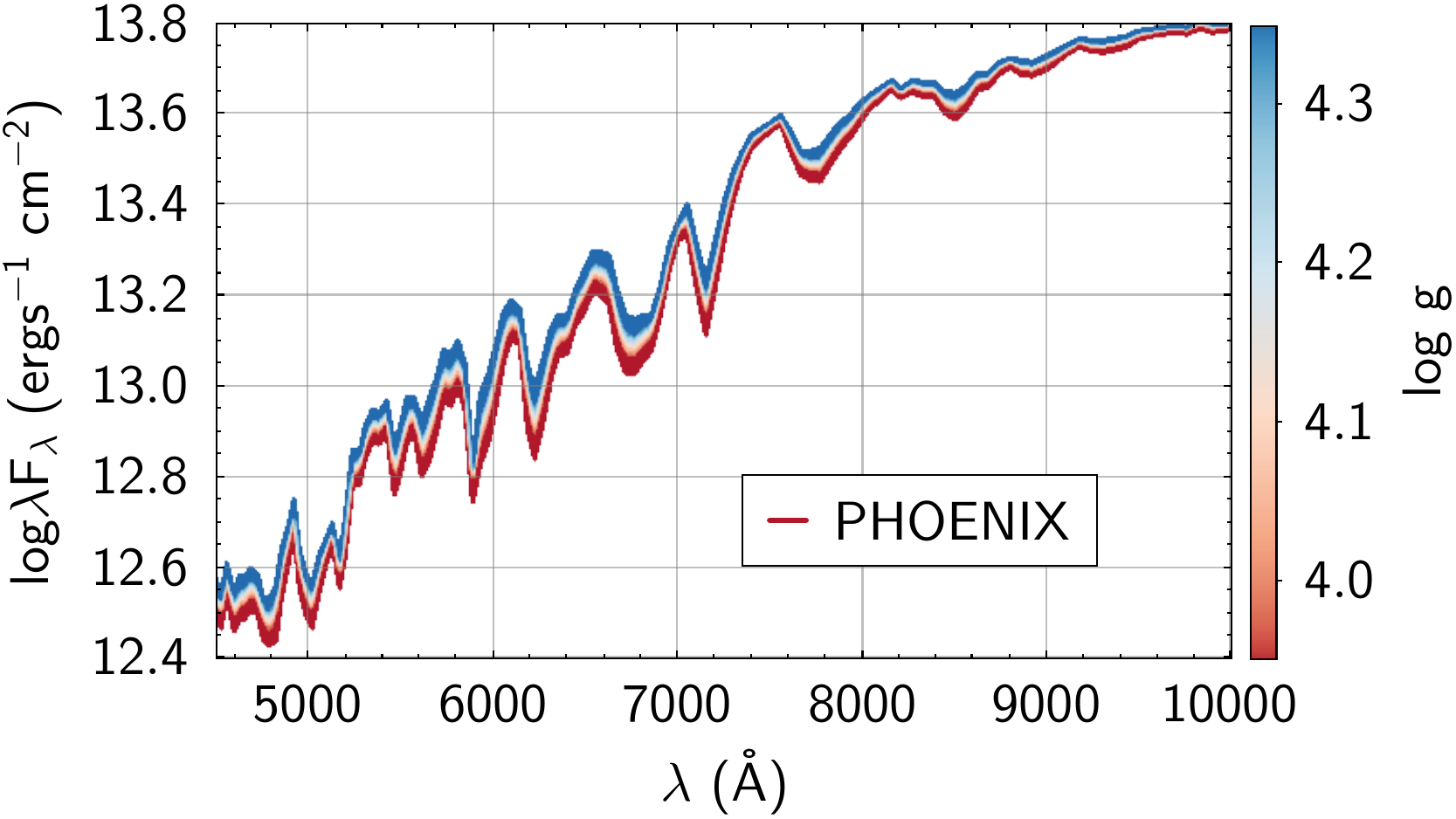}
\caption{Comparison of sampled XP spectra released in Gaia DR3 of pre-main sequence stars (top), versus PHOENIX synthetic spectra \citep[bottom]{husser2013}. XP spectra are selected among the sources with 3450$<$\teff$<$3550 K, and the synthetic spectra has \teff\ of 3500 K. 
\label{fig:loggcomp}}
\end{figure}

In Figure \ref{fig:loggcomp} we compare the spectra of various pre-main sequence stars to identify \logg\ sensitive features. These spectra are selected in a narrow range of \teff s. Through performing SED fitting, we have corrected these spectra for reddening effects, and calculated luminosity per unit surface area (Appendix \ref{sec:sed}). Although \logg\ is generally thought to have a subtle effect, primarily affecting the line broadening that would not be observable at the resolution of XP spectra, there is a considerable role that \logg\ has on the continuum opacity that the model is fundamentally sensitive to. These trends are also seen in synthetic spectra, e.g., PHOENIX \citep{husser2013}, and the overall shape provide a reasonably good match to low resolution XP spectra of pre-main sequence stars in the wavelength range of 6500$<\lambda<$10,000 \AA, even though there is a systematic offset between the depth of the features. The agreement is significantly poorer between 5300$<\lambda<$6500 \AA, where XP spectra plateaus, while the synthetic spectra show broad lines. Possibly this is due to low transmission of XP spectra causing low signal-to-noise \citep{montegriffo2023}. However, given that the synthetic spectra often show a significant discrepancies with the spectra of pre-main sequence stars, this dataset may help refining synthetic models in the future.

\section{Discussion: Discovery of a new young stellar population}

\begin{figure*}
\epsscale{1.15}
        \plottwo{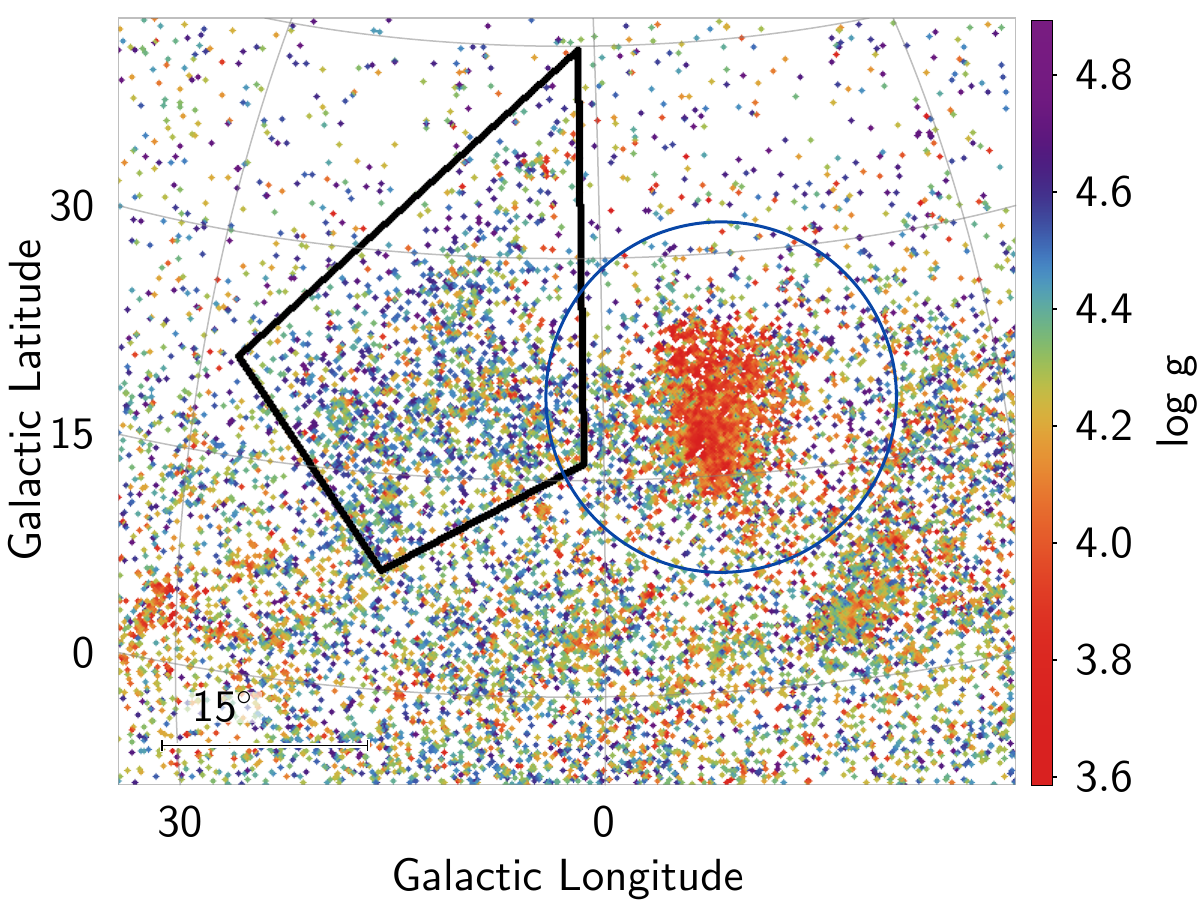}{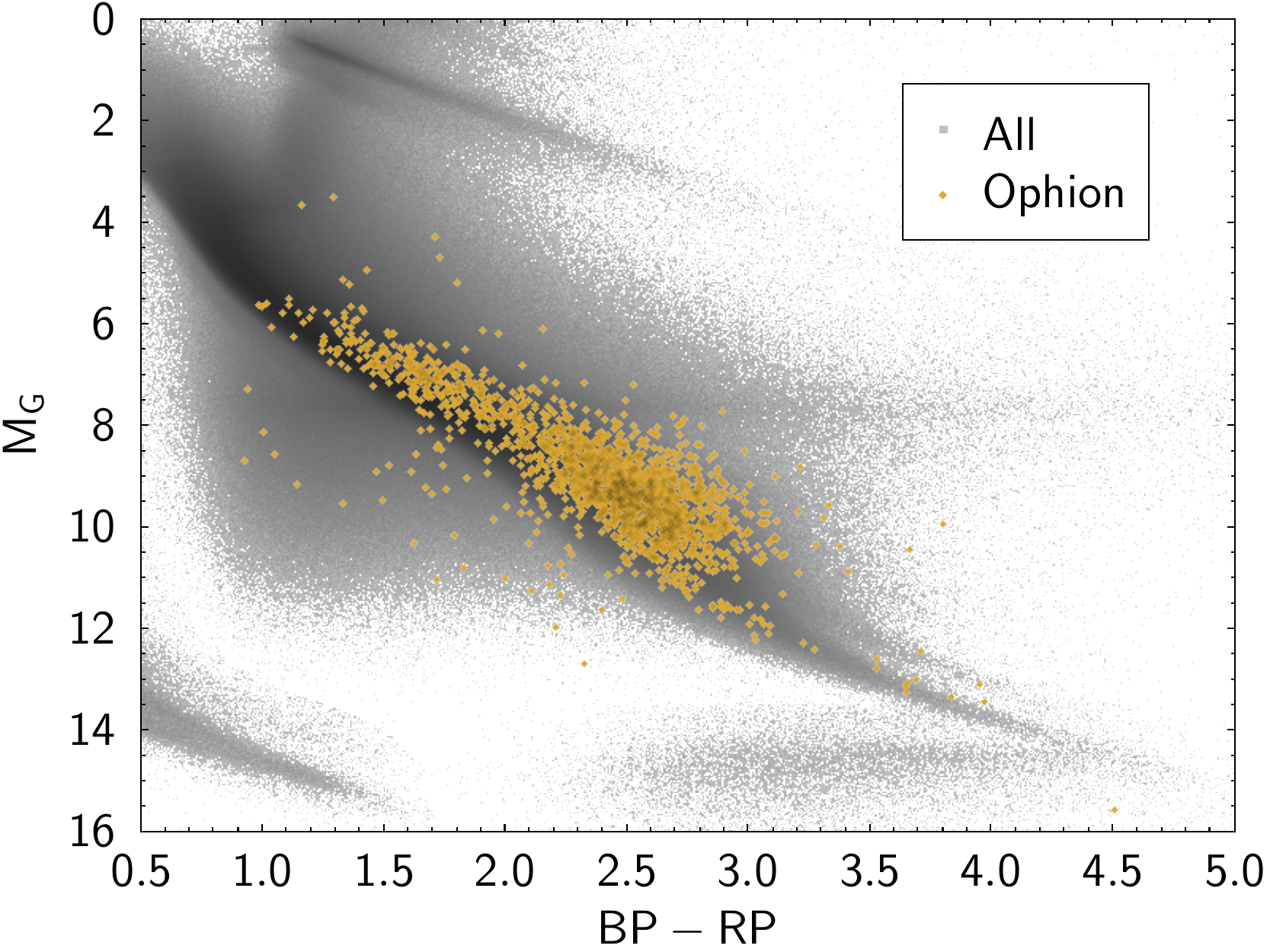}
        \gridline{\leftfig{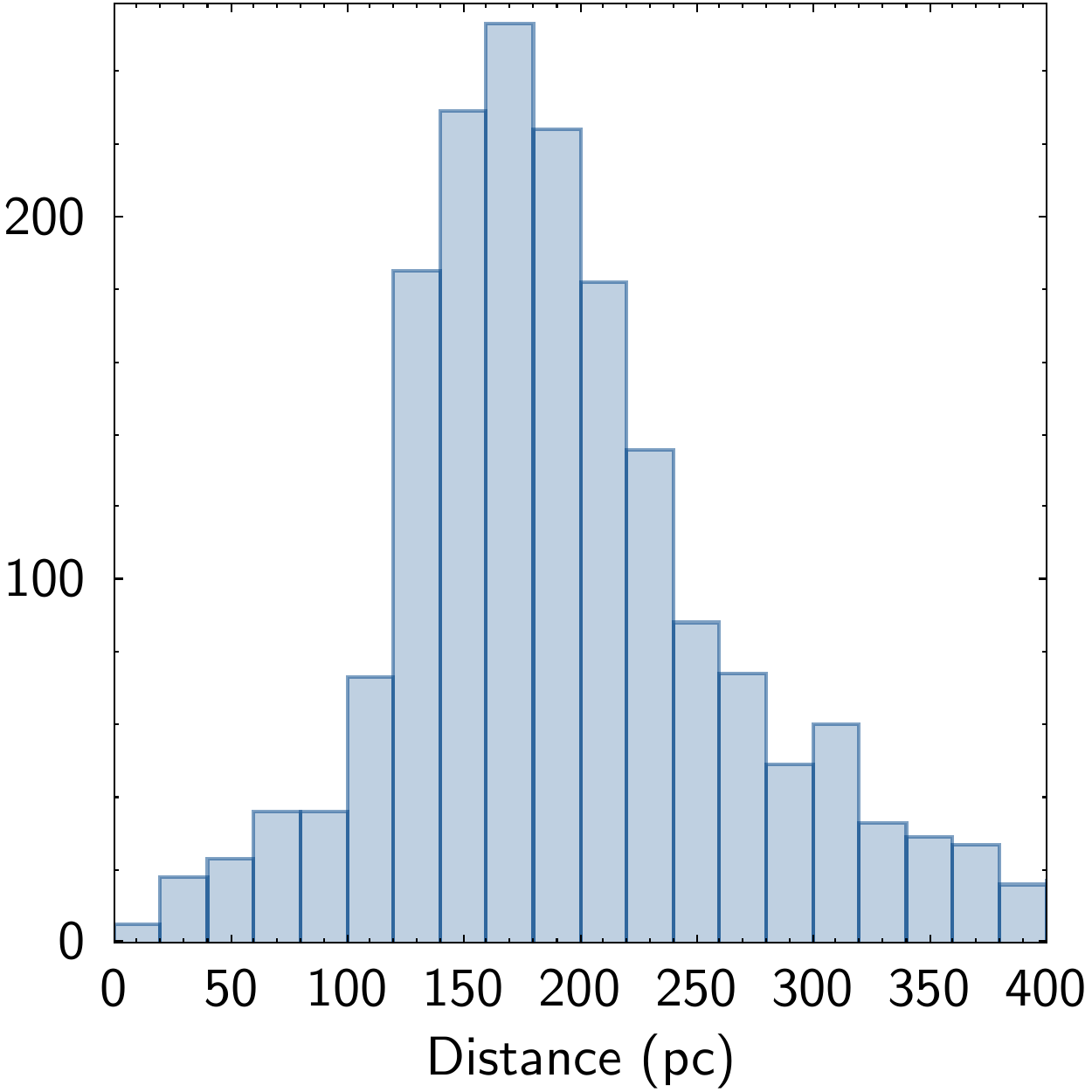}{0.33\textwidth}{}
		          \leftfig{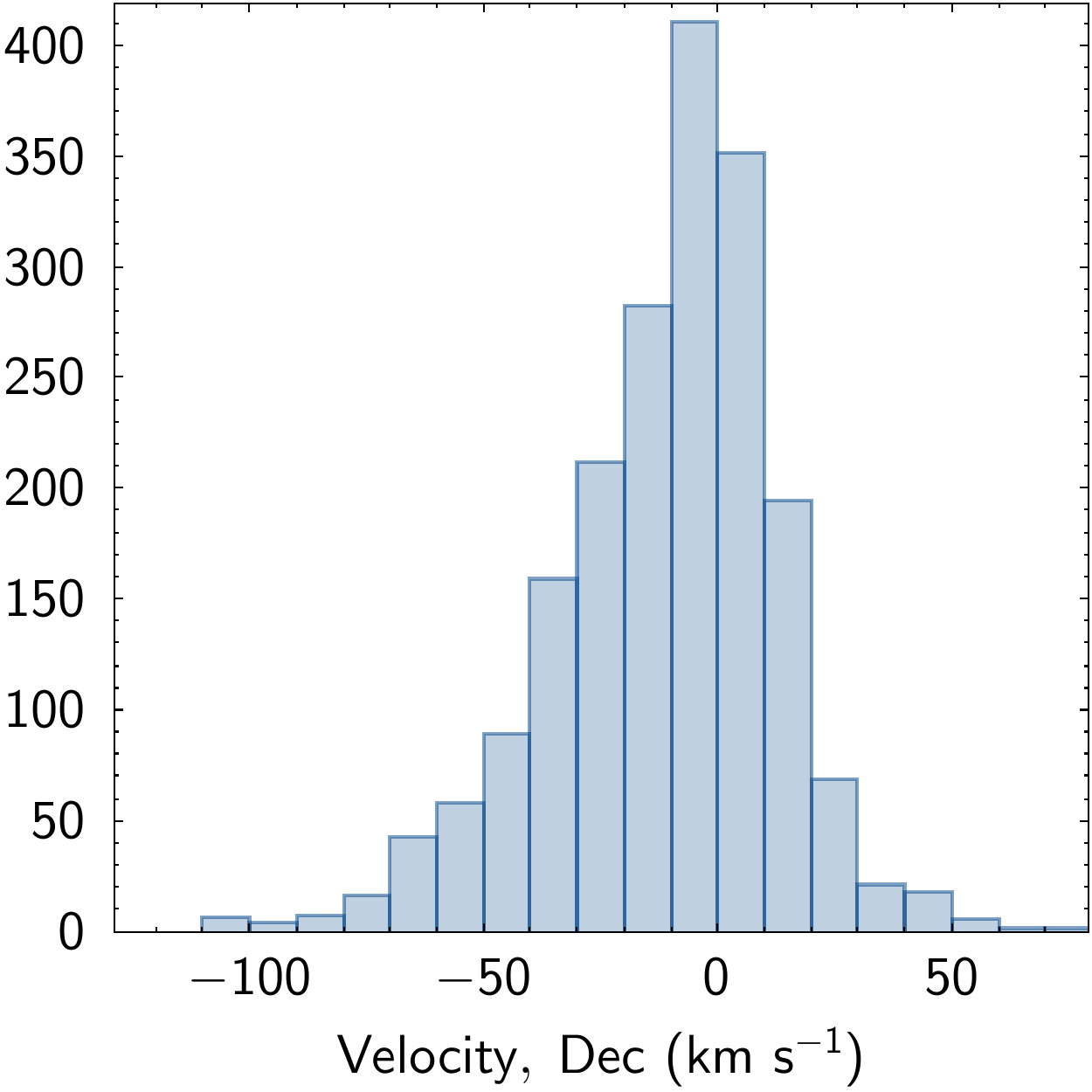}{0.33\textwidth}{}
                \leftfig{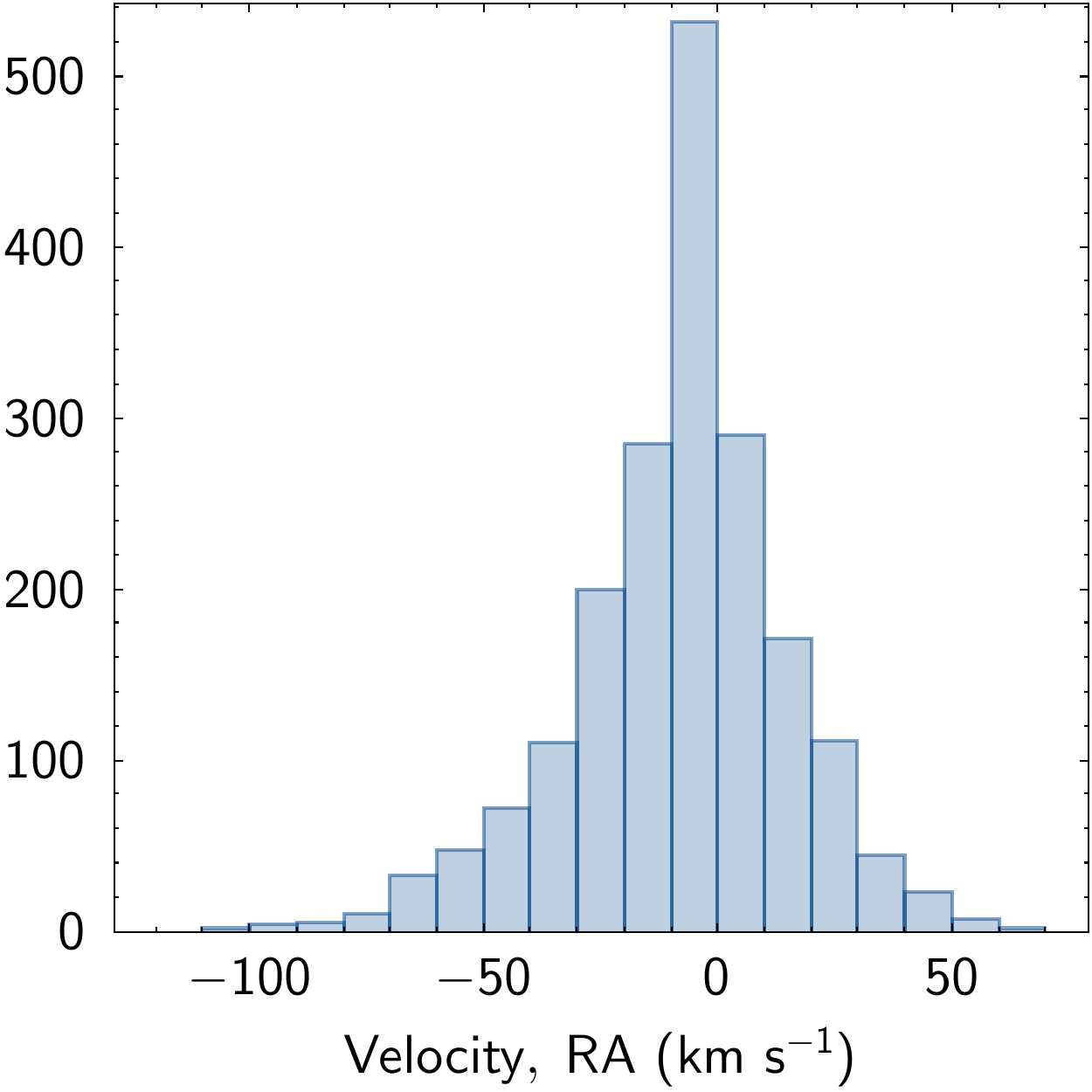}{0.33\textwidth}{}
        }\vspace{-0.8cm}
\caption{Top left: zoom in sky projection towards Ophion, with the overdensity seen towards this region shown by a black outline. The bubble around $\rho$ Oph is shown as a purple circle. Top right: HR diagram of the sources within the black outline, plotted on top of the field stars. Photometery of the field stars is left as is, photometry of Ophion is corrected for $A_V=1.2$ mag, which is typical for the region. Bottom: histogram of distances and on-sky velocities of the sources within the black outline. The proper motions in $\alpha$ and $\delta$ have been converted to the local standard of rest and converted to velocities accounting for the distances (to remove various distortions). None of the projections improve the the coherence of velocities.
\label{fig:oph}}
\end{figure*}

\begin{figure*}
\epsscale{1.15}
        \plotone{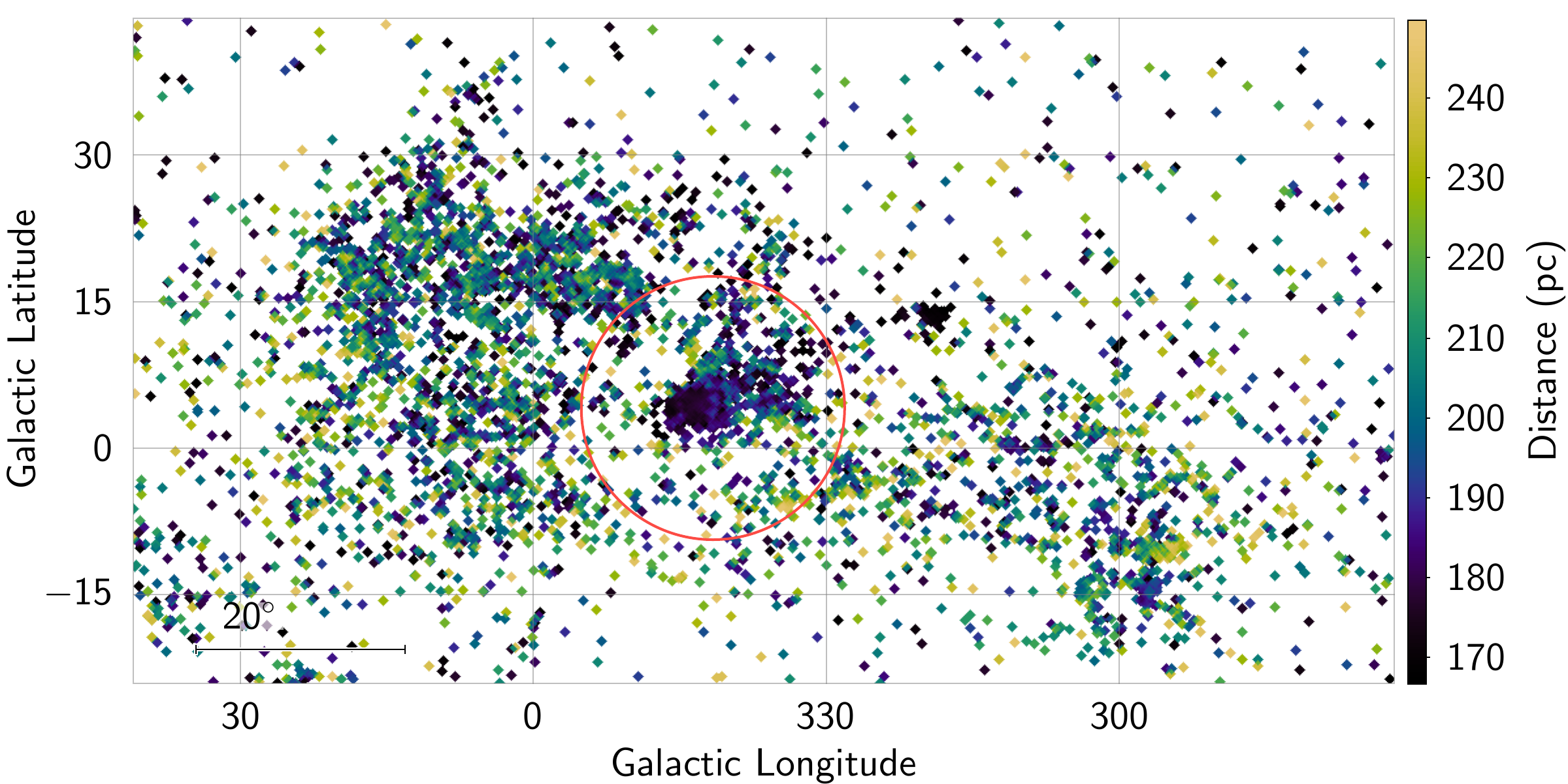}
\caption{Selection of pre-main sequence stars fround between $6>\pi>4$ mas to highlight the full extent of Ophion, featured prominently on the left side of the image. A cluster UPK 640 that is part of Sco Cen is found somewhat on the foreground, and it is seen in the center of the image, the bubble around it is indicated with a red circle. Chameleon clouds are found at a similar distance as Ophion, on the right side of the image.
\label{fig:oph1}}
\end{figure*}

Examining the distribution of pre-main sequence stars produced by the selection of \teff\ and \logg\ does in large part produce a very familiar map of star forming regions that has been frequently observed in several works \citep{zari2018,kounkel2019a,mcbride2021,kerr2021}. However, there does appear to be a somewhat unexpected feature in this map. In particular, east of $\rho$ Oph, there does appear to be a somewhat massive ($>$1000 members) group of stars that appears to have a coherent age of $\sim$20 Myr (clearly pre-main sequence not just on the Kiel diagram, but also on the HR diagram), and concentrated at a distance of $\sim$200 pc (Figure \ref{fig:oph}). Positioned in the middle of the constellation Ophiuchus, to avoid confusion with $\rho$ Oph that is frequently named after this constellation, we refer to this population as Ophion. 

To our knowledge, the only time Ophion was identified previously was as a small clump of 28 stars was found in \citet{ratzenbock2022} labeled as Oph-NorthFar, and it was one of the least significant groups in vicinity of Sco-Cen that was found in that work. As it is located behind a region of moderately high extinction ($\sim1.2$ A$_V$), it generally was not identifiable in photometric searches of pre-main sequence stars. Remarkably, however, despite its number of likely members, Ophion also has almost no kinematic coherence, with the velocity dispersion $>20$ \kms, without any obvious pattern of expansion\footnote{Based on examination of proper motions only. Radial velocities reported by Gaia RVS in DR3 are highly unreliable for pre-main sequence stars \citep{kounkel2023} due to using a bandpass centered on Ca II triplet, which is sensitive to magnetic activity, and require more careful processing. This may be possible to do in future with Gaia DR4. At the moment, no other set of RVs are available for these stars from other surveys.}. This is highly unusual, as star formation occurs in cold molecular clouds with typical velocity dispersion of less than a few \kms, and young stars inherit these kinematics. Star forming regions are rarely gravitationally bound (especially following the gas dispersal) and they do incrementally dissolve in the field as individual stars gain kinetic energy from various tidal kicks. However, it is not unusual to be able to identify moving groups with ages of $>$100 Myr \citep{kounkel2019a}, even if it is just a few dozen stars forming a remnant of what has once been a much more massive population. Thus seeing a massive young population without a coherent ``core'' velocity structure is somewhat unexpected.

At the same time, it cannot be accounted for by selection effects. If \logg\ in this part of the sky are unreliable e.g., due to high extinction, we would expect to find other artifacts in the regions of high extinction, and none are immediately apparent. It also does not trace out the entire extinct region towards its line of sight, which is most strongly concentrated along the Galactic plane and decreases away from it. Similarly, a coherence of the population on the HR diagram, and a coherence in distance also point to Ophion being a real population.

Given its proximity (both in the plane of the sky and in distance), it is likely causally linked to Sco Cen, being an earlier epoch of star formation in this region. It extends beyond the bounds shown in Figure \ref{fig:oph}: examining the sources found at distances of $\sim$200 pc in Figure \ref{fig:oph1} shows that Ophion spreads much further south, to the latitude of CrA. Given that $\rho$ Oph and CrA are both part of Sco Cen \citep{ratzenbock2022}, but Sco Cen is fragmented on the eastern end in the plane of the sky in forming these populations, like a wish-bone, there most likely was an energetic event that has led to this fragmentation. Ophion spans the projected width between these fragments, and it may have been involved in the split, although it is located further behind them. 

There may also be a weak connection between Ophion and the Chameleon clouds (Figure \ref{fig:oph1}), as they are both found at a similar distance, and there appears to be a ``bridge'' between them. 

We further note several apparent bubbles in the distribution of young stars. Figure \ref{fig:oph} shows a bubble centered at $\rho$ Oph, and Figure \ref{fig:oph1} shows a bubble centered at UPK 640. It is possible that these are remnants of supernovae that went off in these regions. Indeed, there have been several voids already identified in vicinity of $\rho$ Oph that have been attributable to supernovae \citep{briceno-morales2022,miret-roig2022}. Supernovae eject molecular gas from the eruption cite in a circular fashion. Subsequently, younger stars form at the edges of the compressed expanding gas, maintaining the shape of the bubble. Additionally, gravitational feedback from the rapid mass loss following gas dispersal can also accelerate already formed stars away from the eruption site \citep{vazquez-semadeni2019}. As such supernovae have significantly impacted star formation in the region.

It is possible the reason for the lack of the kinematical coherence is that Ophion rapidly lost mass during the gas expulsion, more rapidly than what is common in the bulk of the star forming region, possibly through combination of both supernovae and tidal interactions. This resulted in acceleration of the bulk of the stars to very high velocities. And the reason why the region remains coherent spatially is most likely because it is caught in the earliest stages of complete disruption. Given that it's projected size is $\sim$150 pc, both in the plane of the sky and in width, with the velocity dispersion of 20 \kms, with 1 \kms\ translating to $\sim$1 pc Myr$^{-1}$, but with significant structure associated with it still not erased, the loss of velocity support would have likely been recent.

\section{Conclusions}

We have developed a new pipeline, Gaia Net, for reprocessing Gaia XP spectra that is capable of operating across the entire parameter space of \teff\ and \logg, from M\&L dwarfs to O stars, from white dwarfs to supergiants. This pipeline does not require any additional metadata, and makes determination of parameters from spectral coefficients alone. In general, Gaia Net offers improvements for most stars compared to most pipelines that have previously processed XP spectra, although SHBoost \citep{khalatyan2024} has a largely comparable performance.

The area of parameter space where Gaia Net particularly stands out is its performance for pre-main sequence stars, having reliable \logg s that are sensitive to age of young stars, which has only been achieved in pipelines for significantly higher resolution spectra, such as BOSS \& APOGEE \citep{sizemore2024}, as the pipeline is able to make inferences in the global changes in opacity of a star, as opposed to characterizing broadening profiles of the individual lines.

These \logg s offer a new method for examining the star forming history of the solar neighborhood. Instead of determining ages from photometry, which can be biased because of the extinction and multiplicity, \logg\ allows for an independent determination of ages of pre-main sequence stars. Through a preliminary examination of pre-main sequences candidates selected through the use of \logg\ we have been able to identify a new stellar population, Ophion, which is located eastwards of Sco Cen. Ophion is remarkable because despite its age ($<$20 Myr) and mass ($>$1000 stars) it shows very little kinematical coherence, suggesting that the population has recently been fully disrupted. This makes this group to be almost impossible to identify through clustering, which has been a common technique to identify stellar populations to date. In future, through improved determination of ages across the solar neighborhood, it may be possible to identify other such groups which are no longer co-moving to reconstruct them. Combined with improved radial velocity determination for the pre-main sequence stars, which should be possible with Gaia DR4 and SDSS ABYSS \citep{kounkel2023}, it may be possible to shed light on their evolving kinematics. 

\appendix

\section{RVS spectra}\label{sec:rvs}
In addition to reprocessing XP spectra, we have also trained a model with a similar architecture to determine \teff, \logg, and [Fe/H] of the RVS spectra that have been released in Gaia DR3, similarly using the labels from LAMOST and SDSS. Unfortunately, although Gaia DR3 includes radial velocities for almost 34 million stars (therefore, these are the sources for which RVS spectra exists), it only has released the raw spectra for $\sim$1 million stars that are most ``well-behaved objects''. This includes only a handful of pre-main sequence stars, which would be among the sources in the most dire need of reprocessing. This is due to RVS spectra spanning the Ca II triplet, which is often seen in emission in young stars, which is not taken into the account in Gaia DR3 pipeline.

The table of predictions of stellar parameters for the available RVS spectra is shown in Table \ref{tab:rvs}. However, we do note that if the data model of XP spectra in Gaia DR4 is sufficiently similar to what has been released in DR3 it may be possible to apply Gaia Net XP model for all of the sources that would be made available in DR4 as is due to the wide variety of the sources that the model is familiar with, it would unlikely to be the case for the RVS spectra, given the drastic increase in the expected data volume.

\begin{deluxetable}{ccl}[!ht]
\tablecaption{Stellar parameters derived from RVS spectra
\label{tab:rvs}}
\tabletypesize{\scriptsize}
\tablewidth{\linewidth}
\tablehead{
 \colhead{Column} &
 \colhead{Unit} &
 \colhead{Description}
 }
\startdata
source\_id & & Gaia unique identifieer \\
RA & deg & Right ascention in J2000 \\
Dec & deg & Declination in J2000 \\
log \teff & [K] & Effective temperature \\
$\sigma$ log \teff & [K] & uncertainty in log \teff \\
log g &  & Surface gravity \\
$\sigma$ log g & & Uncertainty in log g \\
$\left[\rm{Fe/H}\right]$&  & Metallicity \\
$\sigma$ [Fe/H] & & Uncertainty in [Fe/H] \\
\enddata
\end{deluxetable}

\section{SED}\label{sec:sed}

To examine the intrinsic spectrum of select stars, we have used SEDFit \citep{sedfit}. This code queries broadband UV, optical, and infrared photometery, as well as the sampled Gaia XP spectrum if it has been released by Gaia DR3. It can fit these data with a variety of synthetic spectra, such as, e.g., PHOENIX grid \citep{husser2013}, either solving for \teff\, \logg\, and [Fe/H] using least squares fitting, or through using a single template with the predetermined stellar parameters. As part of the fitting process, the code determines extinction along the line of sight using \citet{gordon2009} profile, and, given parallax, it also determines the radius of a star.

After measuring $A_V$, we then used dust\_extinction package \citep{dustextinction} to divide out the extinction profile from the sampled XP spectra. We further divided out the surface area of the star, assuming spherical geometry, using the determined radius. This enables constructing normalized empirical grid of Gaia XP spectra, and to compare them on the same scale as the synthetic spectra, as can be seen in Figure \ref{fig:loggcomp}.


\software{TOPCAT \citep{topcat}, PyTorch \citep{pytorch}, BOSS Net \citep{sizemore2024}}

\acknowledgments

This work has made use of data from the European Space Agency (ESA)
mission {\it Gaia} (\url{https://www.cosmos.esa.int/gaia}), processed by
the {\it Gaia} Data Processing and Analysis Consortium (DPAC,
\url{https://www.cosmos.esa.int/web/gaia/dpac/consortium}). Funding
for the DPAC has been provided by national institutions, in particular
the institutions participating in the {\it Gaia} Multilateral Agreement.

\bibliography{references.bib, ml_refs.bib}
\bibliography{main.bbl}

\begin{thebibliography}{}
\expandafter\ifx\csname natexlab\endcsname\relax\def\natexlab#1{#1}\fi
\providecommand{\url}[1]{\href{#1}{#1}}
\providecommand{\dodoi}[1]{doi:~\href{http://doi.org/#1}{\nolinkurl{#1}}}
\providecommand{\doeprint}[1]{\href{http://ascl.net/#1}{\nolinkurl{http://ascl.net/#1}}}
\providecommand{\doarXiv}[1]{\href{https://arxiv.org/abs/#1}{\nolinkurl{https://arxiv.org/abs/#1}}}

\bibitem[{{Andrae} {et~al.}(2023){Andrae}, {Fouesneau}, {Sordo},
  {Bailer-Jones}, {Dharmawardena}, {Rybizki}, {De Angeli}, {Lindstr{\o}m},
  {Marshall}, {Drimmel}, {Korn}, {Soubiran}, {Brouillet}, {Casamiquela}, {Rix},
  {Abreu Aramburu}, {{\'A}lvarez}, {Bakker}, {Bellas-Velidis}, {Bijaoui},
  {Brugaletta}, {Burlacu}, {Carballo}, {Chaoul}, {Chiavassa}, {Contursi},
  {Cooper}, {Creevey}, {Dafonte}, {Dapergolas}, {de Laverny}, {Delchambre},
  {Demouchy}, {Edvardsson}, {Fr{\'e}mat}, {Garabato}, {Garc{\'\i}a-Lario},
  {Garc{\'\i}a-Torres}, {Gavel}, {Gomez}, {Gonz{\'a}lez-Santamar{\'\i}a},
  {Hatzidimitriou}, {Heiter}, {Jean-Antoine Piccolo}, {Kontizas}, {Kordopatis},
  {Lanzafame}, {Lebreton}, {Licata}, {Livanou}, {Lobel}, {Lorca}, {Magdaleno
  Romeo}, {Manteiga}, {Marocco}, {Mary}, {Nicolas}, {Ordenovic}, {Pailler},
  {Palicio}, {Pallas-Quintela}, {Panem}, {Pichon}, {Poggio}, {Recio-Blanco},
  {Riclet}, {Robin}, {Santove{\~n}a}, {Sarro}, {Schultheis}, {Segol},
  {Silvelo}, {Slezak}, {Smart}, {S{\"u}veges}, {Th{\'e}venin}, {Torralba
  Elipe}, {Ulla}, {Utrilla}, {Vallenari}, {van Dillen}, {Zhao}, \&
  {Zorec}}]{andrae2023}
{Andrae}, R., {Fouesneau}, M., {Sordo}, R., {et~al.} 2023, \aap, 674, A27,
  \dodoi{10.1051/0004-6361/202243462}

\bibitem[{{Brice{\~n}o-Morales} \& {Chanam{\'e}}(2022)}]{briceno-morales2022}
{Brice{\~n}o-Morales}, G., \& {Chanam{\'e}}, J. 2022, arXiv e-prints,
  arXiv:2205.01735.
\newblock \doarXiv{2205.01735}

\bibitem[{Clevert {et~al.}(2015)Clevert, Unterthiner, \& Hochreiter}]{ELU}
Clevert, D.-A., Unterthiner, T., \& Hochreiter, S. 2015, Under Review of
  ICLR2016 (1997)

\bibitem[{{De Angeli} {et~al.}(2023){De Angeli}, {Weiler}, {Montegriffo},
  {Evans}, {Riello}, {Andrae}, {Carrasco}, {Busso}, {Burgess}, {Cacciari},
  {Davidson}, {Harrison}, {Hodgkin}, {Jordi}, {Osborne}, {Pancino},
  {Altavilla}, {Barstow}, {Bailer-Jones}, {Bellazzini}, {Brown}, {Castellani},
  {Cowell}, {Delchambre}, {De Luise}, {Diener}, {Fabricius}, {Fouesneau},
  {Fr{\'e}mat}, {Gilmore}, {Giuffrida}, {Hambly}, {Hidalgo}, {Holland},
  {Kostrzewa-Rutkowska}, {van Leeuwen}, {Lobel}, {Marinoni}, {Miller},
  {Pagani}, {Palaversa}, {Piersimoni}, {Pulone}, {Ragaini}, {Rainer},
  {Richards}, {Rixon}, {Ruz-Mieres}, {Sanna}, {Sarro}, {Rowell}, {Sordo},
  {Walton}, \& {Yoldas}}]{de-angeli2023}
{De Angeli}, F., {Weiler}, M., {Montegriffo}, P., {et~al.} 2023, \aap, 674, A2,
  \dodoi{10.1051/0004-6361/202243680}

\bibitem[{{Fallows} \& {Sanders}(2024)}]{fallows2024}
{Fallows}, C.~P., \& {Sanders}, J.~L. 2024, \mnras, 531, 2126,
  \dodoi{10.1093/mnras/stae1303}

\bibitem[{{Fouesneau} {et~al.}(2023){Fouesneau}, {Fr{\'e}mat}, {Andrae},
  {Korn}, {Soubiran}, {Kordopatis}, {Vallenari}, {Heiter}, {Creevey}, {Sarro},
  {de Laverny}, {Lanzafame}, {Lobel}, {Sordo}, {Rybizki}, {Slezak},
  {{\'A}lvarez}, {Drimmel}, {Garabato}, {Delchambre}, {Bailer-Jones},
  {Hatzidimitriou}, {Lorca}, {Le Fustec}, {Pailler}, {Mary}, {Robin},
  {Utrilla}, {Abreu Aramburu}, {Bakker}, {Bellas-Velidis}, {Bijaoui}, {Blomme},
  {Bouret}, {Brouillet}, {Brugaletta}, {Burlacu}, {Carballo}, {Casamiquela},
  {Chaoul}, {Chiavassa}, {Contursi}, {Cooper}, {Dafonte}, {Demouchy},
  {Dharmawardena}, {Garc{\'\i}a-Lario}, {Garc{\'\i}a-Torres}, {Gomez},
  {Gonz{\'a}lez-Santamar{\'\i}a}, {Jean-Antoine Piccolo}, {Kontizas},
  {Lebreton}, {Licata}, {Lindstr{\o}m}, {Livanou}, {Magdaleno Romeo},
  {Manteiga}, {Marocco}, {Martayan}, {Marshall}, {Nicolas}, {Ordenovic},
  {Palicio}, {Pallas-Quintela}, {Pichon}, {Poggio}, {Recio-Blanco}, {Riclet},
  {Santove{\~n}a}, {Schultheis}, {Segol}, {Silvelo}, {Smart}, {S{\"u}veges},
  {Th{\'e}venin}, {Torralba Elipe}, {Ulla}, {van Dillen}, {Zhao}, \&
  {Zorec}}]{fouesneau2023}
{Fouesneau}, M., {Fr{\'e}mat}, Y., {Andrae}, R., {et~al.} 2023, \aap, 674, A28,
  \dodoi{10.1051/0004-6361/202243919}

\bibitem[{{Gaia Collaboration} {et~al.}(2018{\natexlab{a}}){Gaia
  Collaboration}, {Brown}, {Vallenari}, {Prusti}, {de Bruijne}, {Babusiaux},
  {Bailer-Jones}, {Biermann}, {Evans}, {Eyer}, \&
  et~al.}]{gaia-collaboration2018}
{Gaia Collaboration}, {Brown}, A.~G.~A., {Vallenari}, A., {et~al.}
  2018{\natexlab{a}}, \aap, 616, A1, \dodoi{10.1051/0004-6361/201833051}

\bibitem[{{Gaia Collaboration} {et~al.}(2018{\natexlab{b}}){Gaia
  Collaboration}, {Babusiaux}, {van Leeuwen}, {Barstow}, {Jordi}, {Vallenari},
  {Bossini}, {Bressan}, {Cantat-Gaudin}, {van Leeuwen}, \&
  et~al.}]{gaia-collaboration2018a}
{Gaia Collaboration}, {Babusiaux}, C., {van Leeuwen}, F., {et~al.}
  2018{\natexlab{b}}, \aap, 616, A10, \dodoi{10.1051/0004-6361/201832843}

\bibitem[{{Gaia Collaboration} {et~al.}(2023){Gaia Collaboration}, {Vallenari},
  {Brown}, {Prusti}, {de Bruijne}, {Arenou}, {Babusiaux}, {Biermann},
  {Creevey}, {Ducourant}, \& et~al.}]{gaia-collaboration2023}
{Gaia Collaboration}, {Vallenari}, A., {Brown}, A.~G.~A., {et~al.} 2023, \aap,
  674, A1, \dodoi{10.1051/0004-6361/202243940}

\bibitem[{Gordon(2024)}]{dustextinction}
Gordon, K. 2024, dust\_extinction, v1.4.1,  Zenodo,
  \dodoi{10.5281/zenodo.11235336}

\bibitem[{{Gordon} {et~al.}(2009){Gordon}, {Cartledge}, \&
  {Clayton}}]{gordon2009}
{Gordon}, K.~D., {Cartledge}, S., \& {Clayton}, G.~C. 2009, \apj, 705, 1320,
  \dodoi{10.1088/0004-637X/705/2/1320}

\bibitem[{He {et~al.}(2016)He, Zhang, Ren, \& Sun}]{he2016deep}
He, K., Zhang, X., Ren, S., \& Sun, J. 2016, in Proceedings of the IEEE
  conference on computer vision and pattern recognition, 770--778

\bibitem[{{Husser} {et~al.}(2013){Husser}, {Wende-von Berg}, {Dreizler},
  {Homeier}, {Reiners}, {Barman}, \& {Hauschildt}}]{husser2013}
{Husser}, T.-O., {Wende-von Berg}, S., {Dreizler}, S., {et~al.} 2013, \aap,
  553, A6, \dodoi{10.1051/0004-6361/201219058}

\bibitem[{Ioffe \& Szegedy(2015)}]{batchnorm}
Ioffe, S., \& Szegedy, C. 2015, CoRR, abs/1502.03167

\bibitem[{{Kerr} {et~al.}(2021){Kerr}, {Rizzuto}, {Kraus}, \&
  {Offner}}]{kerr2021}
{Kerr}, R. M.~P., {Rizzuto}, A.~C., {Kraus}, A.~L., \& {Offner}, S. S.~R. 2021,
  \apj, 917, 23, \dodoi{10.3847/1538-4357/ac0251}

\bibitem[{{Khalatyan} {et~al.}(2024){Khalatyan}, {Anders}, {Chiappini},
  {Queiroz}, {Nepal}, {dal Ponte}, {Jordi}, {Guiglion}, {Valentini}, {Torralba
  Elipe}, {Steinmetz}, {Pantaleoni-Gonz{\'a}lez}, {Malhotra},
  {Jim{\'e}nez-Arranz}, {Enke}, {Casamiquela}, \&
  {Ard{\`e}vol}}]{khalatyan2024}
{Khalatyan}, A., {Anders}, F., {Chiappini}, C., {et~al.} 2024, arXiv e-prints,
  arXiv:2407.06963, \dodoi{10.48550/arXiv.2407.06963}

\bibitem[{Kingma \& Ba(2015)}]{kingma2015}
Kingma, D.~P., \& Ba, J. 2015, in Proceedings of the International Conference
  on Learning Representations.
\newblock \url{http://arxiv.org/abs/1412.6980}

\bibitem[{{Kounkel}(2023)}]{sedfit}
{Kounkel}, M. 2023, mkounkel/SEDFit: 0.3, 0.3,  Zenodo,
  \dodoi{10.5281/zenodo.8076501}

\bibitem[{{Kounkel} \& {Covey}(2019)}]{kounkel2019a}
{Kounkel}, M., \& {Covey}, K. 2019, \aj, 158, 122,
  \dodoi{10.3847/1538-3881/ab339a}

\bibitem[{{Kounkel} {et~al.}(2022){Kounkel}, {Deng}, \&
  {Stassun}}]{kounkel2022}
{Kounkel}, M., {Deng}, T., \& {Stassun}, K.~G. 2022, \aj, 164, 57,
  \dodoi{10.3847/1538-3881/ac7951}

\bibitem[{{Kounkel} {et~al.}(2018){Kounkel}, {Covey}, {Su{\'a}rez},
  {Rom{\'a}n-Z{\'u}{\~n}iga}, {Hernandez}, {Stassun}, {Jaehnig}, {Feigelson},
  {Pe{\~n}a Ram{\'\i}rez}, {Roman-Lopes}, {Da Rio}, {Stringfellow}, {Kim},
  {Borissova}, {Fern{\'a}ndez-Trincado}, {Burgasser},
  {Garc{\'\i}a-Hern{\'a}ndez}, {Zamora}, {Pan}, \& {Nitschelm}}]{kounkel2018a}
{Kounkel}, M., {Covey}, K., {Su{\'a}rez}, G., {et~al.} 2018, \aj, 156, 84,
  \dodoi{10.3847/1538-3881/aad1f1}

\bibitem[{{Kounkel} {et~al.}(2023){Kounkel}, {Zari}, {Covey}, {Tkachenko},
  {Z{\'u}{\~n}iga}, {Stassun}, {Stutz}, {Stringfellow}, {Roman-Lopes},
  {Hern{\'a}ndez}, {Pe{\~n}a Ram{\'\i}rez}, {Bayo}, {Kim}, {Cao}, {Wolk},
  {Kollmeier}, {L{\'o}pez-Valdivia}, \& {Rojas-Ayala}}]{kounkel2023}
{Kounkel}, M., {Zari}, E., {Covey}, K., {et~al.} 2023, \apjs, 266, 10,
  \dodoi{10.3847/1538-4365/acc106}

\bibitem[{{McBride} {et~al.}(2021){McBride}, {Lingg}, {Kounkel}, {Covey}, \&
  {Hutchinson}}]{mcbride2021}
{McBride}, A., {Lingg}, R., {Kounkel}, M., {Covey}, K., \& {Hutchinson}, B.
  2021, \aj, 162, 282, \dodoi{10.3847/1538-3881/ac2432}

\bibitem[{{Miret-Roig} {et~al.}(2022){Miret-Roig}, {Galli}, {Olivares}, {Bouy},
  {Alves}, \& {Barrado}}]{miret-roig2022}
{Miret-Roig}, N., {Galli}, P.~A.~B., {Olivares}, J., {et~al.} 2022, arXiv
  e-prints, arXiv:2209.12938.
\newblock \doarXiv{2209.12938}

\bibitem[{{Montegriffo} {et~al.}(2023){Montegriffo}, {De Angeli}, {Andrae},
  {Riello}, {Pancino}, {Sanna}, {Bellazzini}, {Evans}, {Carrasco}, {Sordo},
  {Busso}, {Cacciari}, {Jordi}, {van Leeuwen}, {Vallenari}, {Altavilla},
  {Barstow}, {Brown}, {Burgess}, {Castellani}, {Cowell}, {Davidson}, {De
  Luise}, {Delchambre}, {Diener}, {Fabricius}, {Fr{\'e}mat}, {Fouesneau},
  {Gilmore}, {Giuffrida}, {Hambly}, {Harrison}, {Hidalgo}, {Hodgkin},
  {Holland}, {Marinoni}, {Osborne}, {Pagani}, {Palaversa}, {Piersimoni},
  {Pulone}, {Ragaini}, {Rainer}, {Richards}, {Rowell}, {Ruz-Mieres}, {Sarro},
  {Walton}, \& {Yoldas}}]{montegriffo2023}
{Montegriffo}, P., {De Angeli}, F., {Andrae}, R., {et~al.} 2023, \aap, 674, A3,
  \dodoi{10.1051/0004-6361/202243880}

\bibitem[{{Olney} {et~al.}(2020){Olney}, {Kounkel}, {Schillinger}, {Scoggins},
  {Yin}, {Howard}, {Covey}, {Hutchinson}, \& {Stassun}}]{olney2020}
{Olney}, R., {Kounkel}, M., {Schillinger}, C., {et~al.} 2020, \aj, 159, 182,
  \dodoi{10.3847/1538-3881/ab7a97}

\bibitem[{Paszke {et~al.}(2017)Paszke, Gross, Chintala, Chanan, Yang, DeVito,
  Lin, Desmaison, Antiga, \& Lerer}]{pytorch}
Paszke, A., Gross, S., Chintala, S., {et~al.} 2017, in NIPS 2017 Workshop on
  Autodiff.
\newblock \url{https://openreview.net/forum?id=BJJsrmfCZ}

\bibitem[{{Ratzenb{\"o}ck} {et~al.}(2023){Ratzenb{\"o}ck}, {Gro{\ss}schedl},
  {M{\"o}ller}, {Alves}, {Bomze}, \& {Meingast}}]{ratzenbock2022}
{Ratzenb{\"o}ck}, S., {Gro{\ss}schedl}, J.~E., {M{\"o}ller}, T., {et~al.} 2023,
  \aap, 677, A59, \dodoi{10.1051/0004-6361/202243690}

\bibitem[{{Recio-Blanco} {et~al.}(2023){Recio-Blanco}, {de Laverny}, {Palicio},
  {Kordopatis}, {{\'A}lvarez}, {Schultheis}, {Contursi}, {Zhao}, {Torralba
  Elipe}, {Ordenovic}, {Manteiga}, {Dafonte}, {Oreshina-Slezak}, {Bijaoui},
  {Fr{\'e}mat}, {Seabroke}, {Pailler}, {Spitoni}, {Poggio}, {Creevey}, {Abreu
  Aramburu}, {Accart}, {Andrae}, {Bailer-Jones}, {Bellas-Velidis}, {Brouillet},
  {Brugaletta}, {Burlacu}, {Carballo}, {Casamiquela}, {Chiavassa}, {Cooper},
  {Dapergolas}, {Delchambre}, {Dharmawardena}, {Drimmel}, {Edvardsson},
  {Fouesneau}, {Garabato}, {Garc{\'\i}a-Lario}, {Garc{\'\i}a-Torres}, {Gavel},
  {Gomez}, {Gonz{\'a}lez-Santamar{\'\i}a}, {Hatzidimitriou}, {Heiter},
  {Jean-Antoine Piccolo}, {Kontizas}, {Korn}, {Lanzafame}, {Lebreton}, {Le
  Fustec}, {Licata}, {Lindstr{\o}m}, {Livanou}, {Lobel}, {Lorca}, {Magdaleno
  Romeo}, {Marocco}, {Marshall}, {Mary}, {Nicolas}, {Pallas-Quintela}, {Panem},
  {Pichon}, {Riclet}, {Robin}, {Rybizki}, {Santove{\~n}a}, {Silvelo}, {Smart},
  {Sarro}, {Sordo}, {Soubiran}, {S{\"u}veges}, {Ulla}, {Vallenari}, {Zorec},
  {Utrilla}, \& {Bakker}}]{recio-blanco2023}
{Recio-Blanco}, A., {de Laverny}, P., {Palicio}, P.~A., {et~al.} 2023, \aap,
  674, A29, \dodoi{10.1051/0004-6361/202243750}

\bibitem[{{Riello} {et~al.}(2021){Riello}, {De Angeli}, {Evans}, {Montegriffo},
  {Carrasco}, {Busso}, {Palaversa}, {Burgess}, {Diener}, {Davidson}, {Rowell},
  {Fabricius}, {Jordi}, {Bellazzini}, {Pancino}, {Harrison}, {Cacciari}, {van
  Leeuwen}, {Hambly}, {Hodgkin}, {Osborne}, {Altavilla}, {Barstow}, {Brown},
  {Castellani}, {Cowell}, {De Luise}, {Gilmore}, {Giuffrida}, {Hidalgo},
  {Holland}, {Marinoni}, {Pagani}, {Piersimoni}, {Pulone}, {Ragaini}, {Rainer},
  {Richards}, {Sanna}, {Walton}, {Weiler}, \& {Yoldas}}]{riello2021}
{Riello}, M., {De Angeli}, F., {Evans}, D.~W., {et~al.} 2021, \aap, 649, A3,
  \dodoi{10.1051/0004-6361/202039587}

\bibitem[{{Rybizki} {et~al.}(2022){Rybizki}, {Green}, {Rix}, {El-Badry},
  {Demleitner}, {Zari}, {Udalski}, {Smart}, \& {Gould}}]{rybizki2022}
{Rybizki}, J., {Green}, G.~M., {Rix}, H.-W., {et~al.} 2022, \mnras, 510, 2597,
  \dodoi{10.1093/mnras/stab3588}

\bibitem[{{Saad} {et~al.}(2024){Saad}, {Lane}, {Kounkel}, {Stassun},
  {L{\'o}pez-Valdivia}, {Kim}, {Pe{\~n}a Ram{\'\i}rez}, {Stringfellow},
  {Rom{\'a}n-Z{\'u}{\~n}iga}, {Hern{\'a}ndez}, {Wolk}, \&
  {Hillenbrand}}]{saad2024}
{Saad}, S., {Lane}, K., {Kounkel}, M., {et~al.} 2024, \aj, 167, 125,
  \dodoi{10.3847/1538-3881/ad2001}

\bibitem[{{Sizemore} {et~al.}(2024){Sizemore}, {Llanes}, {Kounkel},
  {Hutchinson}, {Stassun}, \& {Chandra}}]{sizemore2024}
{Sizemore}, L., {Llanes}, D., {Kounkel}, M., {et~al.} 2024, arXiv e-prints,
  arXiv:2402.05184.
\newblock \doarXiv{2402.05184}

\bibitem[{{Sprague} {et~al.}(2022){Sprague}, {Culhane}, {Kounkel}, {Olney},
  {Covey}, {Hutchinson}, {Lingg}, {Stassun}, {Rom{\'a}n-Z{\'u}{\~n}iga},
  {Roman-Lopes}, {Nidever}, {Beaton}, {Borissova}, {Stutz}, {Stringfellow},
  {Ram{\'\i}rez}, {Ram{\'\i}rez-Preciado}, {Hern{\'a}ndez}, {Kim}, \&
  {Lane}}]{sprague2022}
{Sprague}, D., {Culhane}, C., {Kounkel}, M., {et~al.} 2022, \aj, 163, 152,
  \dodoi{10.3847/1538-3881/ac4de7}

\bibitem[{{Storey-Fisher} {et~al.}(2024){Storey-Fisher}, {Hogg}, {Rix},
  {Eilers}, {Fabbian}, {Blanton}, \& {Alonso}}]{storey-fisher2024}
{Storey-Fisher}, K., {Hogg}, D.~W., {Rix}, H.-W., {et~al.} 2024, \apj, 964, 69,
  \dodoi{10.3847/1538-4357/ad1328}

\bibitem[{{Szil{\'a}gyi} {et~al.}(2023){Szil{\'a}gyi}, {Kun},
  {{\'A}brah{\'a}m}, \& {Marton}}]{szilagyi2023}
{Szil{\'a}gyi}, M., {Kun}, M., {{\'A}brah{\'a}m}, P., \& {Marton}, G. 2023,
  arXiv e-prints, arXiv:2301.02346.
\newblock \doarXiv{2301.02346}

\bibitem[{{Taylor}(2005)}]{topcat}
{Taylor}, M.~B. 2005, in Astronomical Society of the Pacific Conference Series,
  Vol. 347, Astronomical Data Analysis Software and Systems XIV, ed.
  P.~{Shopbell}, M.~{Britton}, \& R.~{Ebert}, 29

\bibitem[{{V{\'a}zquez-Semadeni} {et~al.}(2019){V{\'a}zquez-Semadeni}, {Palau},
  {Ballesteros-Paredes}, {G{\'o}mez}, \&
  {Zamora-Avil{\'e}s}}]{vazquez-semadeni2019}
{V{\'a}zquez-Semadeni}, E., {Palau}, A., {Ballesteros-Paredes}, J.,
  {G{\'o}mez}, G.~C., \& {Zamora-Avil{\'e}s}, M. 2019, \mnras, 490, 3061,
  \dodoi{10.1093/mnras/stz2736}

\bibitem[{{Zari} {et~al.}(2018){Zari}, {Hashemi}, {Brown}, {Jardine}, \& {de
  Zeeuw}}]{zari2018}
{Zari}, E., {Hashemi}, H., {Brown}, A.~G.~A., {Jardine}, K., \& {de Zeeuw},
  P.~T. 2018, \aap, 620, A172, \dodoi{10.1051/0004-6361/201834150}

\bibitem[{{Zhang} {et~al.}(2023){Zhang}, {Green}, \& {Rix}}]{zhang2023}
{Zhang}, X., {Green}, G.~M., \& {Rix}, H.-W. 2023, \mnras, 524, 1855,
  \dodoi{10.1093/mnras/stad1941}

\end{thebibliography}

\end{document}